# The Property Law of Crypto Tokens

Jakub Wyczik[*]

Abstract

This article addresses the lack of comprehensive studies on Web3 technologies, primarily due to lawyers' reluctance to explore technical intricacies. Understanding the underlying technological foundations is crucial to enhance the credibility of legal opinions. This article aims to illuminate these foundations, debunk myths, and concentrate on determining the legal status of crypto-assets in the context of property rights within the distributed economy. In addition, this article notes that the intangible nature of crypto-assets that derive value from distributed registries, and their resistance to deletion, makes crypto-assets more akin to the autonomy of intellectual property than physical media. The article presents illustrative examples from common law (United States, United Kingdom, New Zealand) and civil law (Germany, Austria, Poland) systems. Proposing a universal solution, it advocates a comprehensive framework safeguarding digital property—data ownership—extending beyond the confines of Web3.

---

[*] Institute of Law, University of Silesia in Katowice, Poland; jakub.wyczik@us.edu.pl, https://www.linkedin.com/in/jakub-wyczik/, https://orcid.org/0000-0003-4169-4670. This research was funded in whole or in part by National Science Centre, Poland 2021/41/N/HS5/02726. For the purpose of Open Access, the author has applied a CC-BY public copyright license to any Author Accepted Manuscript (AAM) version arising from this submission. No potential competing interest was reported by the author. This manuscript is licensed under CC BY 4.0. To view a copy of this license, visit http://creativecommons.org/licenses/by/4.0/

## 1. Introduction

This article presents the issue of the private law status of tokens, focusing on the ownership of tokens as the broadest exclusive right, effective against all persons.

First, I will thoroughly explain the technical aspects of tokens. Adopting appropriate definitions and adequately determining the facts remain crucial for conducting a legal analysis of any social relations. It is necessary to clearly emphasize the multi-layered structure of digital assets and the multidimensionality of the associated legal relations. In my opinion, there are at least three types of token rights: (1) ownership of the token as a virtual good, (2) rights to the assets that can be linked to the token, and (3) rights to the intellectual property (IP) embedded in or otherwise associated with the token.

Next, I present the well-established perception of tokens as property in common law systems, particularly in England and Wales, New Zealand, and Singapore. I then analyze the concept of token rights prevalent in civil law countries, describing solutions in Poland, Germany, and Austria, among others. Equally important is the section devoted to rights linked to tokens, such as claims. This is because tokens can entitle their holders to receive certain benefits from other entities. In this way, they can be similar to certificates or vouchers. The last part of this article focuses on the issue of intellectual property rights (IPRs) over intangible assets embedded in or otherwise associated with tokens.

I can summarize the considerations of this article in one sentence: the need to recognize the concept of digital data as objects of property rights is a recurring problem that is becoming more important every year. Today, we can say that we have reached a critical point. The lack of a legislative initiative to regulate the digital data at this stage may lead to property rights becoming a relic of the past. We live at a time when more and more goods are offered as services rather than products. With continued attempts to justify the legal qualification of such services solely as contract claims, people will cease to be owners of the goods they purchase. Instead, the world will be merely a network of complex contractual relationships between parties. From this stage, the world will be one step away from declaring the independence of the metaverse controlled by the tech giants and having its autonomy recognized by its users. In this scenario, tokens are just the tip of the iceberg.

It is, therefore, necessary to develop solutions that treat digital data as things and, thus, as objects of property rights in both common law and civil law systems. On the other hand, viewing property rights solely through the prism of existing constructs, such as the requirement to exercise physical control over an object, is unjustified. Thus, even in common law systems, we should carefully distinguish between the right to possess an object and ownership vested in a particular person regardless of who possesses the object at a given time. This would solve the problem of tokens and other virtual goods or digital content. However, for such legal solutions to be possible and fully effective, states must cooperate, and outdated (pre-Internet) legal concepts need to be changed. The same can be said of IP, particularly the economically unjustified application of the first sale doctrine exclusively to tangible copies of a work. This is despite the fact that this form of distribution of works is now a thing of the past.

## 2. Understanding technology and property

### 2.1. Concept of property

In analyzing the right to own tokens, it is first necessary to outline what such a right is supposed to consist of. Ownership is often presented as a triad of rights derived in the continental system from Roman law: (1) the right to possess the thing, (2) the right to use the thing, and (3) the right to dispose of the thing. Without going into historical details, I will focus on two available approaches: the civil law system, especially from the perspective of Polish and German law, and the general concept of property in common law systems.

I must clarify at the start that the English language uses the term property, which is ambiguous. In principle, it can mean both the object that belongs to someone and the right itself to the object. The object (*res*) itself is also referred to as 'thing' or 'chose,' although these terms also refer to traditional ownership. Therefore, the distinction between property rights and their objects is blurred. The literature generally uses the term property to refer to rights to the possession and use of a given good.[1] Sometimes, ownership is distinguished from property as the state of being the owner of a particular good or property interest, which characterizes the person who exercises fundamental control over the thing.[2] Such ambiguity and other differences between the civil law and common law systems make it impossible to maintain uniform terminology, although the two systems have much in common.

---

[1] Bryan A Garner and Henry Campbell Black (eds), *Black's Law Dictionary* (Eleventh, Thomson Reuters 2019); Andreas Rahmatian, *Lord Kames: Legal and Social Theorist* (Edinburgh University Press 2016) 221–24.

There is no need to discuss all the details about these systems. However, as is well known, one of the backbones of property is possession, which common law systems treat as a right, and the person with the best title to possession is considered the owner.[3] Thus, common law already protects possession, which it does through remedies for torts, such as trespass to goods and conversion. On the other hand, civil law systems generally protect the right to property through special petition claims, which include a claim for the retribution of a thing (*rei vindicatio*) and a negatory claim for the cessation of other violations and restoration of the lawful state of affairs (*actio negatoria*).[4] Possession in civil law is a matter of fact rather than a right to control a good. In both systems, however, possession is based on two elements: (1) the exercise of actual control (referred to in civil law as *corpus*) and (2) the intention to exercise exclusive control over a specific good (*animus*).[5] Due to the common law link between possession and property rights, loss of possession is generally considered a loss of property right. This can occur, for example, through the transfer of possession to another person or the deliberate relinquishment of possession by abandoning control.[6] In civil law, property right over an asset can generally continue to exist despite the loss of control over the asset.

---

[2] Lutz-Christian Wolff, 'The Relationship between Contract Law and Property Law' (2020) 49 Common Law World Review 1, 31, 35.

[3] Andreas Rahmatian, 'A Comparison of German Moveable Property Law and English Personal Property Law' (2008) 3 Journal of Comparative Law 1, 197, 211–14.

[4] ibid 214–16.

[5] ibid 207.

[6] ibid 209.

Both systems distinguish between tangible and intangible goods, but in the common law, both categories are parts of the concept of property. Civil law generally grants individuals exclusive rights to intangible goods without recognizing such goods as objects of property rights.[7] Despite some differences between the two systems, they appear to remain convergent in their scope on many issues. However, these differences in practice seem to raise significant challenges in addressing the issue of exclusive rights to intangible assets, such as tokens.

Nevertheless, the ownership of any property is about to exclude others from the good in question because of the effort required to obtain it. Classically, such appropriation is possible through physical actions. However, if only physicality were the key, whoever has more strength or cunning could take the good from someone else. What about IPRs or claims? Neither of these rights directly targets physical objects. A narrowly defined physical action does not provide sufficient protection in these cases. Therefore, the need arose for adequate legal protection, which we know today as property rights, an independently protected value that guarantees the ability to use goods.

Regardless of the legal system, property rights are merely fiction, in which we recognize a special relationship of a certain person to a particular good, which gives them the right to control it. However, property rights exist because the law recognizes and treats certain goods as protected objects. The objects affected by this right are merely the abstract legal concept of thing (*res*, thing, chose).[8]

---

[7] ibid 201.

[8] Andreas Rahmatian, 'Debts, Money, Intellectual Property, Data and the Concept of Dematerialised Property' (2020) 11 JIPITEC 2, 186, 188.

## 2.2. Taxonomy of tokens

On the other hand, there is the concept of tokens. Generally, tokens have different meanings depending on the sector and context.[9] For example, a token can be an electronic device for generating authentication codes.[10] In this article, however, when I speak of tokens, I mean tokens operating within networks based on distributed ledger technology (DLT).[11]

DLT is primarily a paradigm for managing data within distributed databases, which generally do not have a central administrator. The technology is used to create networks centered around specific rules aimed primarily at achieving consensus in replicating, sharing, and synchronizing geographically dispersed data across multiple locations.[12] Thus, looking at the most popular DLT-based networks, one could say they are 'networks inside the Internet.' They might provide space (hosting) for creating and operating further solutions. Such technologies, as a rule, are supported by advanced cryptographic solutions, which explains the emerging suffix crypto for related issues.

To understand tokens, it is best to verify developers' understanding of such solutions. OpenZeppelin's documentation defines a token as 'a representation of something in the blockchain. This something can be money, time, services, shares in a company, a virtual pet, anything. By representing things as tokens, we can allow smart contracts[13] to interact with them, exchange them, create or destroy them.'[14] This definition is ideologically true, although we must emphasize that blockchain is only one of many ways to implement DLT. Tokens can also exist as a part of DLT implementations other than blockchains, such as the hashgraph used within the Hedera Hashgraph network.[15] More precisely, we can say that a token is simply a particular digital asset, digital content, or virtual good, functioning on a DLT-based network. In this way, a token is the equivalent of a thing in Web3.[16]

---

[9] Claudia Di Bernardino and others, 'NFT – Legal Token Classification' (European Union Blockchain Observatory and Forum 21 July 2021) 2.

[10] 'Security Token', *Wikipedia* (2022).

[11] 'Distributed Ledger', *Wikipedia* (2022).

[12] Mark Walport, 'Distributed Ledger Technology: Beyond Block Chain' (Government Office for Science 2016).

[13] A type of computer program. See Part A.II.2.

Strictly speaking, we should also distinguish between a token and a cryptocurrency. A cryptocurrency is a native medium of exchange for a given DLT network (for example, bitcoin in the Bitcoin network, ether in Ethereum, or HBAR in Hedera).[17] Sometimes, however, this distinction can be elusive, as, depending on the context, both terms might refer interchangeably to the same assets. It is generally accepted that a cryptocurrency is also any digital currency that serves as a medium of exchange independent of any central authority (for example, a bank or a government) that is supposed to have control over it and is needed for its operation.[18] For example, ether is a natural currency in Ethereum. It is also called 'gas' or 'fuel' because it is used to pay for the computing resources of a distributed virtual machine—Ethereum.[19] Leaving the details aside, tokens are undoubtedly digital data stored within the framework of the mechanisms mentioned above. As a rule, they are specific assets that are cryptographically secured. Therefore, they are collectively referred to as crypto-assets.

---

[14] 'Tokens - OpenZeppelin Docs' (*OpenZeppelin*) <https://docs.openzeppelin.com/contracts/3.x/tokens> accessed 23 December 2022.

[15] 'How It Works' (*Hedera*) <https://hedera.com/how-it-works> accessed 23 December 2022.

[16] Web3 is the idea of a new iteration of the Internet using DLT and the concepts of decentralization and token-based economics. Gavin Wood, co-founder of Ethereum, is considered its creator, presenting the above vision in 2014. See Gavin Wood, 'What Is Web 3? Here's How Future Polkadot Founder Gavin Wood Explained It in 2014' (4 January 2022) <https://www.coindesk.com/layer2/2022/01/04/what-is-web-3-heres-how-future-polkadot-founder-gavin-wood-explained-it-in-2014/> accessed 23 December 2022.

[17] Vladimir, 'Coin vs. Token, Explanation Of The Difference Between Coins And Tokens - Zerion' (*Zerion Blog*, 28 July 2022) <https://zerion.io/blog/crypto-coins-vs-tokens-the-difference-explained/> accessed 23 December 2022.

[18] Claus Dierksmeier and Peter Seele, 'Cryptocurrencies and Business Ethics' (2018) 152 Journal of Business Ethics 1, 1.

### 2.2.1. Token classification

Since a token is a certain asset, it can arguably be classified. As with most assets, tokens can be subdivided limitlessly, depending on the criteria adopted. However, they are most fundamentally divided based on fungibility into the following:

- Fungible tokens (FTs) – which are exchangeable for other assets with essentially the same value, such as ether or bitcoin (similar to fiat currencies); and
- Non-fungible tokens (NFTs) – the opposite of FTs, as they are not substitutable and have individual properties.

The above divisions may seem familiar to lawyers. The criterion for this is the substitutability of assets. However, this division may be ambiguous as the tokens' essence is the same. Depending on the context, the same asset can be fungible or non-fungible. The assessment depends on the parties' will, declared explicitly or implicitly.[20] How the object is defined (for example, the degree of precision and separation) is decisive.[21] Therefore, it is important to remember that a token representing one share in the decentralized autonomous organization (DAO)[22] or a token representing one share in the DAO but with the identification number 666 can be traded exactly the same as a hundred-dollar bill or a hundred-dollar bill with the serial number AA 00000001.

---

[19] 'Gas and Fees' (*ethereum.org*) <https://ethereum.org/en/developers/docs/gas/> accessed 23 December 2022.

[20] Royston Miles Goode and Ewan McKendrick, Goode and McKendrick on Commercial Law (Sixth, Penguin Books 2020) para 2.89. 'Fungibility is thus fundamentally dependent on the particular obligation, usually contractual, rather than inherent to any property, tangible or otherwise.' See also Roy Goode, 'Are Intangible Assets Fungible?' (2003) 3 Lloyd's Maritime and Commercial Law Quarterly 3, 379, 383; David Fox, Property Rights in Money (Oxford University Press 2008) paras 1.78-1.81.

However, such a division may be a matter of objective considerations—for example, real estate always represents non-fungible thing. Objective considerations in the case of tokens are primarily technical standards. If the tokens are created in the FT standard (for example, ERC-20),[23] they are not numbered (i.e. they are not unique and cannot be individually specified). Thus, the transaction can rely only on an object specified in quantity, which means that the tokens are interchangeable. If the tokens are numbered (for example, ERC-721 standard),[24] the object of sale can be a token with ID one but not two or three. This means the transaction is about this particular token (i.e. an NFT).

---

[21] Sąd Najwyższy [Polish Supreme Court], 18 December 1973, I CR 363/73.

[22] DAOs are organizations that operate based on rules encoded in the form of smart contracts. They are governed by their members without centralized leadership. See 'Decentralized Autonomous Organization', *Wikipedia* (2022).

[23] 'ERC-20 Token Standard' (*ethereum.org*) <https://ethereum.org/en/developers/docs/standards/tokens/erc-20/> accessed 23 December 2022.

[24] 'ERC-721 Non-Fungible Token Standard' (*ethereum.org*) <https://ethereum.org/en/developers/docs/standards/tokens/erc-721/> accessed 23 December 2022.

The resources on crypto-assets distinguish yet another category of tokens: the so-called 'semi-fungible tokens' (SFTs).[25] It is said that they can change their status. For example, an event ticket with an individual number remains fungible with other tickets of the same type until the event occurs. After the event, the ticket remains a non-fungible good with a collector's value identified by a specific serial number. This undoubtedly highlights the essence of fungibility, which must be evaluated on a case-by-case basis, depending on the parties' will and the circumstances. However, SFTs are more of a term for a certain smart contract standard that allows FTs and NFTs to coexist within a single program (for example, ERC-1155).[26] Thus, this category does not seem relevant for the legal assessment.

The aforementioned division can therefore be presented in such a way that in the case of FTs, the question is *how many*, and in the case of NFTs, *which*. NFTs are thus often referred to as 'unique' and 'indivisible.'[27] However, such an approach is imprecise. Uniqueness is often viewed literally and taken to mean that each NFT is one of a kind. Meanwhile, because of how DLT networks work, creating a new smart contract, even on the same network, that will store a token with attributes similar to those of an earlier one will not be an issue. There simply cannot be two tokens with the same ID within a smart contract. But this does not mean that there cannot be two entries in other smart contracts that point to the same image. Just as there can be two digital files that store the same content.

---

[25] Anatol Antonovici, 'What Is a "Semi-Fungible" Crypto Token?' (17 August 2021) <https://www.coindesk.com/tech/2021/08/17/what-is-a-semi-fungible-crypto-token/> accessed 23 December 2022.

[26] 'ERC-1155 Multi-Token Standard' (*ethereum.org*) <https://ethereum.org/pl/developers/docs/standards/tokens/erc-1155/> accessed 23 December 2022.

[27] Di Bernardino and others (n 9) 2.

In practice, we refer to uniqueness more regarding the standards mentioned above. Two NFTs can exist like two cars of the same brand, with the same model, color, and configuration but different vehicle identification numbers. Each of these cars is unique and one of a kind, but they are not the same. Thus, the point is that there is no same token, although there may be a similar token.

Tokens can be also classified in other ways, such as by their function,[28] as follows:

- Utility tokens – treated like vouchers and other mediums of exchange;
- Security or equity tokens – often perceived as stocks, bonds, or other securities; and
- Governance tokens – which represent control over the governance of specific projects, in particular the ability to vote on specific issues, such as the future of a DAO.

The above classification is relevant in practice for the transfer of further rights. This is because the sale of a token does not always lead to the buyer taking over all of the seller's rights (for example, under the copyright or additional privileges). On the other hand, the token might be seen almost as a kind of certificate if, for example, the issuer considers that each holder of a particular token will receive a specific benefit. There are many possibilites, but they all depend on individual circumstances to the extent that a token may not even guarantee any additional rights.

As a rule, a token is an independent asset and can be considered without associated rights. But it seems that the buyer will want to acquire the full rights arising from the typical purpose of the good associated with the token. Then, as a rule, the greater value of such data is usually only apparent when combined with further rights. Creating such a bond, however, requires an appropriate approach to creating and applying regulations to reflect the public's intentions and protect market participants. This article focuses on NFTs as the most developed tokens. However, the solutions presented are universal and apply to other tokens.

---

[28] The literature offers several ways to formulate such divisions. Rafał Skibicki and others, 'Algorithmisation and Tokenisation of Law' in Dariusz Szostek and Mariusz Załucki (eds), *Legal Tech* (Nomos Verlagsgesellschaft mbH & Co KG 2021) 45.

### 2.2.2. The basics of tokens and the role of smart contracts

NFTs are instinctively associated with digital art, images, animations, and other audiovisual content. This seems natural but is technologically oversimplified. To properly understand what tokens are, it is necessary to explain how they work.

First, it is categorically essential to distinguish between a token and a smart contract. These terms are often confused and used synonymously, but they refer to different goods. To explain the problem, I will use the example of Ethereum. Tokens on the Ethereum network are created based on smart contracts, which are computer programs placed in a database that operates within the said network. Smart contracts used to manage tokens are called 'token contracts.' They enable the creation, storage, and disposal of specific tokens. Token users transfer tokens to other people by invoking a particular subroutine stored in a smart contract that someone previously programmed. This subroutine is supposed to assign the tokens, when executed, to a new user due to a change of records in the database. Such a smart contract is thus basically nothing more than a set of instructions leading to the creation of a database of addresses assigned to specific numbers. These numbers correspond to the tokens. Smart contracts have several additional functions that are primarily responsible for performing operations on this data (for example, checking the balance of specific addresses, viewing files associated with tokens, or verifying the holders of specific tokens). NFTs are stored in the form of unique identifiers to which the addresses of their holders are assigned. On the other hand,

FTs are merely numerical values assigned to the aforementioned addresses.[29] In technical terms, a token is an aforementioned number representing a specific asset. In some simplifications, such a database can be represented as follows.

*Table 1*

*Simplified representation of token records*

For fungible tokens:

| Address | Balance |
|---|---|
| 0x0000…0000 | 0 |
| 0x11f2…ae24 | 3500 |
| 0x56s3…18e2 | 1200 |
| 0x93f1…c028 | 20 |
| 0xe99b…3492 | 489 |

For non-fungible tokens:

| Address | Token ID |
|---|---|
| 0x0000…0000 | 1 |
| 0x11f2…ae24 | 2 |
| 0x56s3…18e2 | 3 |
| 0x93f1…c028 | 4 |
| 0xe99b…3492 | 5 |

Operations on FTs involve simple mathematical operations, such as subtracting a certain value from the existing holder's balance and increasing the purchaser's balance. On the other hand, with NFTs, the address of the holder assigned to a given token ID is changed. Once again, the real uniqueness of NFTs lies in the non-duplicability of the ID numbers within a single smart contract.

To fully illustrate the basics of tokens, explaining the aforementioned addresses is necessary. Networks such as Ethereum operate based on the so-called 'accounts' of users, which form sets of the following three data:

1) Private key – essentially any value that the user chooses. There are often no specific requirements as to how this value should be chosen, although it is recommended that it be as random as possible to minimize the likelihood of another user generating an existing account;

---

[29] Jim McDonald, 'Understanding ERC-20 Token Contracts | Weald Technology' (24 April 2019) <https://www.wealdtech.com/articles/understanding-erc20-token-contracts/> accessed 23 December 2022.

2) Public key – a value generated from the private key by applying a specific algorithm (for example, in Ethereum, the Elliptic Curve Digital Signature Algorithm); and

3) Address – the value created after applying a specific algorithm on the public key (for example, in Ethereum, the last 20 bytes of the Keccak-256 public key hash preceded by '0x').[30]

The essence of account is that the above values are transformed into successive values using advanced one-sided cryptography. This means that a person with a private key can generate a public key and address, but not the other way around.[31] Therefore, the private key must remain private, according to its name. It is equivalent to a password. On the other hand, the public key can be compared to a login name, and the address, to a user name. The difference between Ethereum accounts and standard access data is that the private key determines other values, which makes the private key the most important part of the entire account.

At the same time, it is worth noting that in the Ethereum network, smart contracts also have their accounts and, therefore, their addresses. However, they do not have a private key, so a smart contract cannot initiate a transaction. Nevertheless, the smart contract has an owner, who can, for example, withdraw the funds accumulated for the tokens sold. Thus, smart contracts are often described as the equivalent of vending machines.[32] However, smart contracts that act like token stores generally do not give out tokens to the purchasers but assign their names to the tokens and store the tokens on their behalf.

---

[30] Hash means the result of the encryption function.

[31] 'Ethereum Accounts' (*ethereum.org*) <https://ethereum.org/en/developers/docs/accounts/> accessed 23 December 2022.

[32] See 'Nick Szabo -- Smart Contracts: Building Blocks for Digital Markets' (1996) <https://www.fon.hum.uva.nl/rob/Courses/InformationInSpeech/CDROM/Literature/LOTwinterschool2006/szabo.best.vwh.net/smart_contracts_2.html> accessed 23 December 2022.

To make a transaction on tokens, one must undergo an authorization process. It involves generating, through an appropriate algorithm, a kind of signature based on the private key, the content of the transaction, and additional variables. Another algorithm verifies that the signature correctly represents the signer's address as it appears in the body of the transaction message. If the data match, the network nodes consider the transaction valid and add it to the distributed ledger.[33]

As with tokens and smart contracts, an account and its components (i.e. the private key, public key, and address) should not be confused with a wallet. A wallet is just a product, such as software or hardware, that facilitates account management. Therefore, it is accurate to compare a wallet with an online banking application without a bank.[34] The wallet allows one to check their balance, make transactions, and connect to apps, but it is not the account itself, in the same way that a bank's mobile app is not a bank account. The wallet only stores and manages accounts, acting as a password manager.

### 2.2.3. Multi-layered structure of the token

After explaining the essential technical background, the construction of tokens must be considered. Tokens have a multi-layered nature. NFTs are typically identified by various graphics assigned to individual tokens. These graphics are economically unviable to store on many DLT-based networks, such as Ethereum, because of the high fees for storing large files in the databases of smart contracts. Therefore, in most cases, the smart contract's database stores only a hash that hides a link to a metadata file in which individual token attributes are specified. These attributes typically contain a further hash to an image or other representation. Someone may use a centralized server to store these files, so the file may no longer be available after some time. In general, however, the idea is to preserve the principle of decentralization of assets. To achieve that, solutions such as the InterPlanetary File System (IPFS) are used, which provides 'decentralized' file hosting based on a caching mechanism.[35] But even such a system can fail at some point. Then, the expected graphics or other content will no longer appear on the user's device. Fortunately, 100% on-chain projects, which involve uploading entire files on a DLT network, are becoming increasingly popular. To store data in some way in distributed networks, one can also use a form of encoding, such as base64, which stores data in binary form.[36]

---

[33] Network nodes are the devices on which the network software runs. They can be said to be the equivalent of servers that verify all transactions, thus ensuring the security of the network. 'Nodes and Clients' (*ethereum.org*) <https://ethereum.org/en/developers/docs/nodes-and-clients/> accessed 23 December 2022.

[34] 'Ethereum Wallets' (*ethereum.org*) <https://ethereum.org/en/wallets/> accessed 23 December 2022.

To illustrate the above dependencies, I will use the example of a well-known token that represents the graphic EVERYDAYS: THE FIRST 5000 DAYS by Mike Winkelmann (known professionally as 'Beeple').[37] The token was sold through Christie's auction house for about $69 million in ether. The token has been embedded in a smart contract on the Ethereum

---

[35] 'What is IPFS? | IPFS Docs' (*ipfs.tech*) <https://docs.ipfs.tech/concepts/what-is-ipfs/> accessed 23 December 2022.

[36] See mbvissers.eth, 'How to Implement Fully On-Chain NFT Contracts' (*Quick Programming*, 11 February 2022) <https://medium.com/quick-programming/how-to-implement-fully-on-chain-nft-contracts-8c409acc98b7> accessed 23 December 2022.

[37] 'Beeple (b. 1981), EVERYDAYS: THE FIRST 5000 DAYS | Christie's' <https://onlineonly.christies.com/s/beeple-first-5000-days/beeple-b-1981-1/112924> accessed 23 December 2022.

network, so anyone can check its metadata because of its public availability. After calling the tokenURI function, the program returns an IPFS hash to the JSON file shown in Figure 1.

*Figure 1*

*Beeple's Art NFT – JSON File ('Metadata')*

![JSON metadata screenshot showing fields: title, name, type, imageUrl, description, attributes (trait_type: Creator, value: beeple), properties including name, description, preview_media_file, preview_media_file_type (jpg), created_at (2021-02-16T00:07:31.674688+00:00), total_supply (1), digital_media_signature_type (SHA-256), digital_media_signature (6314b55cc6ff34f67a18e1ccc977234b803f7a5497b94f1f994ac9d1b896a017), raw_media_file]

*This content is not covered by the terms of the Creative Commons licence of this publication. For permission to reuse, please contact the rights holder.*

The JSON file's contents define the token's attributes, specifying its name, type, and appearance. Subsequent hashes refer to image files representing the appearance of the token, which are stored with IPFS (Figure 2).

*Figure 2*

*Beeple's Art NFT – Image File*

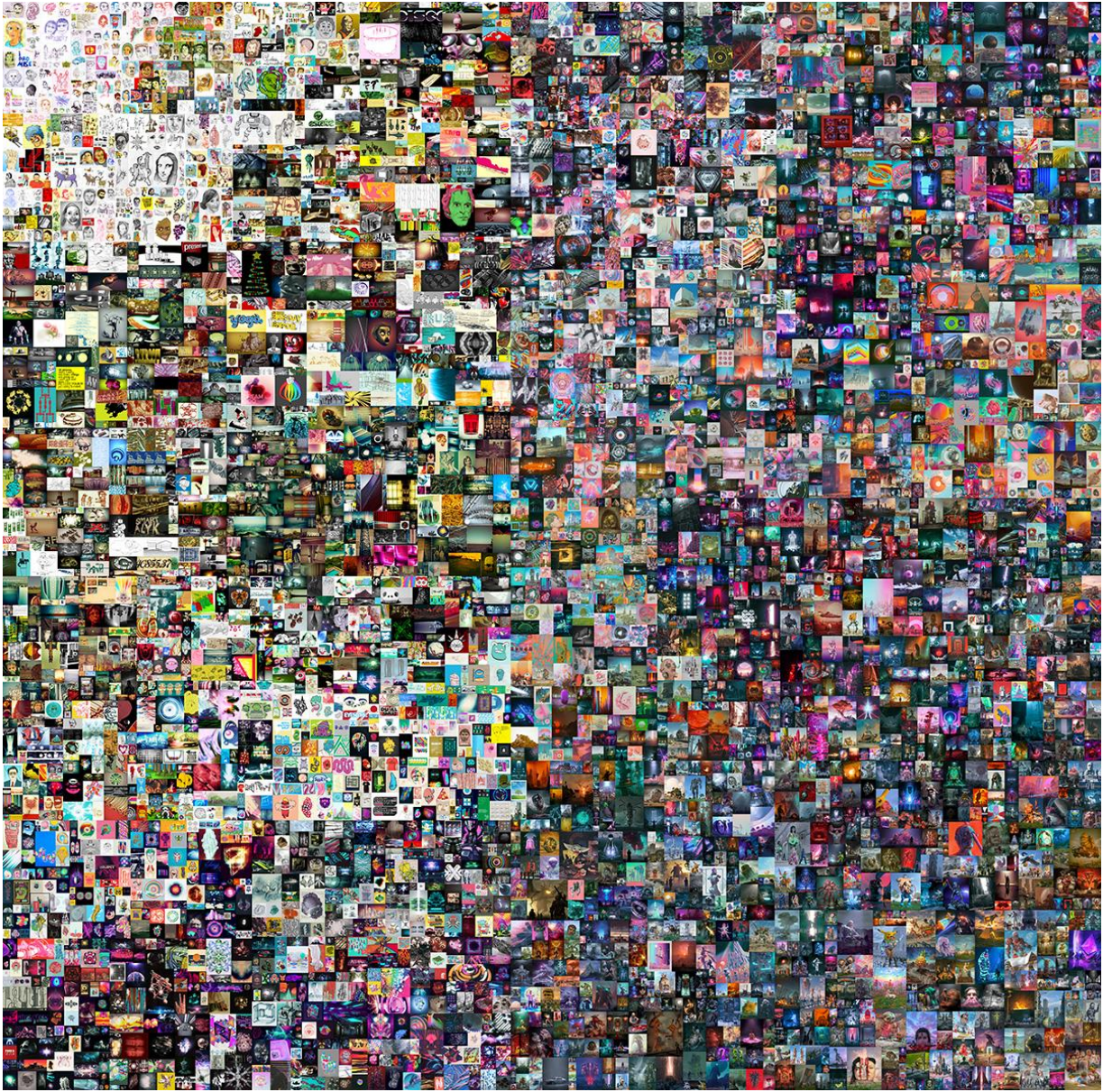

This example shows only one possible solution, although it is undoubtedly one of the most popular. Files may also be stored in other ways, notably via another DLT network, such as LBRY, on which the Odysee platform operates.[38] Moreover, the token does not have to refer to any content and can even be described solely through a single attribute, such as a 'name.' However, it has become standard for NFTs to have a graphical representation.

The above example illustrates that, in principle, NFTs, understood holistically as digital assets, consist mainly of at least two layers:

1) Digital layer – contains digital records of the tokens in the smart contract's database and their corresponding values and attributes, including their metadata; and

2) Media layer – represents the desired effect of executing certain computer commands, which usually results in generating the selected image, sound, and other content on the user's device; inevitably connected with tokens but need not be the subject of uniform token trading.

However, from a technical viewpoint, tokens are generally merely data digitally stored in a distributed ledger that can be traded.[39] The NFT is just a unique identifier associated with the address of its purported owner, which is stored at the address of a particular smart contract. Depending on the circumstances, this identifier may be mapped with something more, which is a generally desirable feature of NFT, but not inherent.

---

[38] See 'LBRY - Content Freedom' (*LBRY*) <https://lbry.com/> accessed 23 December 2022.

[39] See Kathleen Bridget Wilson and others, 'Prospecting Non-Fungible Tokens in the Digital Economy: Stakeholders and Ecosystem, Risk and Opportunity' (2022) 65 Business Horizons 5, 657; Christina Majaski, 'Cryptocurrency Security Token: Definition, Forms, Investing In' (*Investopedia*, 23 December 2022) <https://www.investopedia.com/terms/s/security-token.asp> accessed 23 December 2022.

## 3. Multi-layered property in Web3

The construction of tokens naturally affects the legal qualification of property rights. It is necessary to distinguish between the right to dispose of a token and the right to dispose of the various goods associated with it. Therefore, we can say that in Web3, as in the material world, the property is multidimensional and cannot be reduced to the ownership of a single thing. In practice, we can consider the rights to various goods in the token trading market divisible into at least the following three categories:

1) Rights to the token as a virtual good;

2) Rights to assets that can be linked to the token; and

3) Rights to the IP embedded in the token or otherwise associated with it.

At the same time, we must remember that additional issues require legal evaluation, including aspects of token creation. In particular, the right of communication to the public and smart contracts' author rights. The main problem lies in the questionable grounds for protecting the authors of works from the use of their IP to create tokens. Undoubtedly, uploading graphics or other files containing works on a hosting platform can be considered at least one reproduction of the work. However, a fundamental question arises as to whether the mere act of linking can be counted as an infringement of the author's rights when the alleged infringer uses a file already uploaded to the Internet with the right holder's permission.[40] Linking a file containing a work to a token can be as simple as placing a link in the metadata. Possible direct copyright infringement could therefore consist solely of communication to the public.[41] A detailed discussion of these issues is beyond the scope of this article.[42]

---

[40] See Case C-392/19 *VG Bild-Kunst v Stiftung Preußischer Kulturbesitz* [2021] ECLI:EU:C:2021:181; Case C-161/17 *Land Nordrhein-Westfalen v Dirk Renckhoff* [2018] ECLI:EU:C:2018:634; Case C-160/15 *GS Media BV v Sanoma Media Netherlands BV* [2016] ECLI:EU:C:2016:644.

### 4. Rights to the token as a virtual good

The rights to a token as a virtual good raise the fundamental question on the existence of virtual property. The answer varies by jurisdiction. Because of their inherent flexibility, common law countries can better withstand technological innovations. The following section answers the above question from the perspective of the existing case law.

#### 4.1. Common law

##### 4.1.1. England and Wales, New Zealand, and Singapore

The possibility of seeing crypto-assets as objects of property rights has been noted directly in common law countries based on English precedents.[43] One of the most frequently cited decisions is the December 2019 English judgment *AA v Persons Unknown, Re Bitcoin*.[44] At the time, an English court issued an injunction over the 96 units of bitcoin cryptocurrency that hackers stole.

---

[41] See *TuneIn Inc v Warner Music UK Ltd and Sony Music Entertainment UK Ltd* [2021] EWCA Civ 441 (Eng); *American Broadcasting Cos Inc v Aereo Inc* 573 US 431 (2014). However, the mere visual incorporation of a work by embedding does not constitute display under the 'server test' set forth in *Perfect 10 Inc v AmazonCom Inc* 508 F3d 1146 (9th Cir 2007).

[42] See Andres Guadamuz, 'The Treachery of Images: Non-Fungible Tokens and Copyright' (2021) 16 Journal of Intellectual Property Law & Practice 12, 1367.

[43] *Vorotyntseva v MONEY-4 Ltd (t/a nebeus.com)* [2018] EWHC 2596 (Ch) [13] (holding that there was no suggestion that cryptocurrency could not be a form of property and that the court could not prevent a party from dealing in it).

[44] *AA v Persons Unknown, Re Bitcoin* [2019] EWHC 3556 (Comm).

The case involved a ransomware victim whose ransom was paid in bitcoins. However, it was possible to trace a specific address on one of the exchanges where part of the ransom was still located. In light of this, the victim applied for an injunction. To effectively prevent the disposition of an asset, one must qualify it as a subject of property rights. Classically, in English law, ownership of personal property is divided into two categories: choses in possession and choses in action.[45] However, according to the court, cryptocurrencies are problematic because they do not qualify as choses in possession.[46] This is because they are virtual by their very nature; therefore, as intangible objects, they cannot be in someone's physical possession.[47] They also do not qualify as choses in action because they do not embody any right that could be asserted solely by bringing an action in court.[48] These statements cause difficulty because in *Colonial Bank v Whinney*, the court said that all personal things constitute one of the above categories, and the law knows no third type of property.[49]

However, referring to the UK Jurisdictional Task Force's Legal Statement on Cryptoassets and Smart Contracts published in November 2019,[50] the court found that such a statement is incorrect.[51] Sir Simon Bryan cited after the above report that the *Colonial Bank v Whinney* case was strictly about bankruptcy law and did not resolve the scope of property rights.[52] In addition, lawyers at the time inaccurately used the division to identify different categories of property, even suggesting the need to extend choses in action to all intangibles.[53] A recent case on the above observation involved a pledge on a digital database.[54] Sir Martin Moore-Brick pointed out that there is a strong argument for reconsidering the dichotomous distinction between choses in action and choses in possession and for recognizing the existence of a third category of intangible property that may also be subject to peculiar possession.[55] However, the court could not decide on the case as it was bound by the *OBG Ltd v Allan* decision.[56]

---

[45] *Colonial Bank v Whinney* (1885) 30 ChD 261 [285].

[46] *AA* (n 44) [55].

[47] See *Armstrong DLW GmbH v Winnington Networks Ltd* [2012] EWHC 10 (Ch) [44].

[48] *Torkington v Magee* [1902] 2 KB 427.

[49] *AA* (n 44) [55].

[50] UK Jurisdicitonal Task Force, 'Legal Statement on Cryptoassets and Smart Contracts' (UK Jurisdicitonal Task Force 2019).

[51] *AA* (n 44) [56]-[58].

The authors of the UK Jurisdictional Task Force report emphasized that in *Your Response Ltd v Datateam Business Media Ltd*, the court only answered whether the good could be subject to certain legal remedies. According to the authors, even if the question of ownership should have been examined, the case involved a database that could not be considered property because it contained pure information.[57] However, such a statement is incorrect.

One must note the difference between information and digital data. In principle, a database can be viewed in terms of both.[58] A database as information is about the protection of an intangible asset in the form of knowledge regarding certain objects that, occurring in a given context, acquire a specific meaning. A database, understood as digital data, is an actual representation of that information. Thus, when talking about the ownership of a database as information, we generally talk about IPRs, which protect the producer's effort to collect and organize the information. For a database as digital content, we talk about ownership of the dataset, an information medium used to store a copy of the database (for example, a computer file). Just as we distinguish the author's rights to a written novel from the printed sheets, we should also distinguish a database from its digital copies. Of course, it is quite easy to duplicate the number of copies of digital content. However, many copies are also created in the computer RAM due to the legal use of such copies. Adopting the above assumptions, therefore, requires considering a certain legal fiction, which assumes that the ownership of digital data applies in principle to the 'same' copy of a certain asset and not to the 'original' copy.[59] However, this does not prevent such goods from being viewed in terms of property. Similarly, the Supreme Court of New Zealand in *Dixon v R* decided that a digital copy of a camera recording could be a form of property.[60]

---

[52] See *Allgemeine Versicherungs-Gesellschaft Helvetia v Administrator of German Property* [1931] 1 KB 672.

[53] UK Jurisdicitonal Task Force (n 50) 76.

[54] *Your Response Ltd v Datateam Business Media Ltd* [2014] EWCA 281 (Civ).

[55] ibid [27].

[56] *OBG Ltd v Allan* [2007] UKHL 21. Interestingly, Lord Hoffman stated (para 101) that a domain name may be intangible property, such as a copyright or trademark, but it does not seem reasonable to attribute liability for the tort of conversion to the registrar of such property.

[57] UK Jurisdicitonal Task Force (n 50) 81.

[58] See the definitions of data and information in ISO/IEC 2382:2015.

Returning to the consideration of the cryptocurrency case, one may find that in many 20th-century legal acts, English law uses a broad conception of property as things in action 'and other intangible property.'[61] Because of the arguments given in the report above, Sir Simon Bryan held that cryptocurrencies, such as bitcoins, constitute property within the meaning of English law. This is because, according to the court, they meet all four criteria outlined in the classic definition of property by Lord Wilberforce in *National Provincial Bank v Ainsworth*.[62] Nevertheless, the case lacked a detailed analysis of the above criteria.

---

[59] See Case C-128/11 *UsedSoft GmbH v Oracle International Corp* [2012] ECLI:EU:C:2012:407.

[60] *Dixon v R [2015] NZSC 147*; cf Anusha Wijewickrama, 'Dixon v R – Property in Digital Information?' (2017) 7 VUWLRP 74/2017, 1, 16–33. However, the court did not specify how the scope of such a right would look like. See *Henderson v Walker* [2019] NZHC 2184 (holding that the tort of conversion also includes interference with another's property in the form of digital information such as the contents of private email correspondence, with the mere copying of such information not yet constituting the tort but only the refusal of access to it or its deletion).

However, the above ruling, while probably the most popular, was not the first case to involve crypto-assets in the context of property.[63] This reference was also in the March 2019 ruling of the Singapore International Commercial Court.[64] The case involved a dispute between a cryptocurrency exchange and a user trading using automated algorithms. A platform error allowed the user to exchange ether for bitcoins at a rate 250 times lower than the market rate. However, shortly after the transactions were discovered, the operator unilaterally canceled them and restored previous balances. The user sued the platform operator for breach of contract and trust. The court found that cryptocurrencies met all the above criteria for being considered property and could therefore be the subject of a trust. However, since the dispute did not directly concern ownership, the court did not consider the matter further.[65] Nevertheless, the claim based on the breach of trust was dismissed at the appeal stage due to the lack of certainty about the intent to create the trust.[66]

---

[61] Eg UK Jurisdictional Task Force pointed to the Theft Act 1968, the Proceeds of Crime Act 2002, and the Fraud Act 2006, and also pointed out that the Patents Act 1977 goes even further by stating that a patent or patent application 'is personal property (without being a thing in action).'

[62] *AA* (n 44) [59].

[63] Although most of the cases actually referred to cryptocurrencies, the arguments and conclusions contained in the judgments can be applied by analogy to other types of crypto-assets.

[64] *Quoine Pte Ltd v B2C2 Ltd* [2019] SGHC(I) 03.

[65] ibid [142]; *Quoine Pte Ltd v B2C2 Ltd* [2020] SGCA(I) 02 [137]-[144].

More detailed considerations can be found in the High Court of New Zealand decision *Ruscoe v Cryptopia Ltd (In Liquidation)*.[67] The dispute again concerned the existence of trust between Cryptopia, a New Zealand cryptocurrency exchange, and its users. Cryptopia was hacked, and lost part of the funds stored in cryptocurrency accounts. As a result, the company had to declare bankruptcy. Users asked the liquidators to return the rest of their funds. The liquidators asked the court for guidance on the legal status of these funds, including the possibility of considering them property under the Companies Act 1993. If the funds were the property of the users, the platform would not be able to satisfy the claims of all creditors. The court thus analyzed in detail all the criteria for the classic definition of the property presented in *National Provincial Bank v Ainsworth*.

1) Identifiable subject matter

The first requirement is that the asset must be separable from other assets.[68] In the case of cryptocurrencies, the court found that this condition is met because funds are stored in machine-readable strings. Therefore, it is possible to assign specific funds to an account holder in a particular network. The assignment is made via an address and a public key so that data assigned to one key will not be confused with others. This claim is reinforced by using cryptography, which helps ensure stability. It is irrelevant that identical data are stored on each computer connected to the network.[69] The court compared how cryptocurrencies work to what happens in the banking system, where trusted international banks record balances on various account numbers held in those banks. Cryptocurrencies stored on distributed networks are as identifiable as bank account balances. Equity treats recorded bank balances as the property of the person in whose name the balance is recorded.[70]

---

[66] *Quoine Pte Ltd* [2020] (n 65) [145]-[149].

[67] *Ruscoe v Cryptopia Ltd (In Liquidation)* [2020] NZHC 728.

[68] ibid [104].

[69] ibid [105].

2) Identifiable by third parties

The second requirement relates to the ability of third parties to identify the owner of the asset. Thus, there must be sufficient control over a particular asset. The primary concern is the potential realization of the owner's right to exclude others from the asset.[71] Again, according to the court, cryptocurrencies meet the above conditions. First, the unique data values that record transactions are always assigned by address to a specific user of a specific account. However, this is not sufficient. In the case of cryptocurrencies, there is a corresponding degree of control through a mechanism that assigns to each address an additional set of data in the form of a public key and a private key. The private key grants the account holder the exclusive right to decide on the transfer of the cryptocurrency associated with the corresponding public key.[72] According to the court, the above exclusion of others is reinforced by the general generation of a new set of keys after each transfer.[73] The court also said that this protects against transferring the same unit of cryptocurrency twice. However, this observation is not correct, just as changing the password to a bank account does not protect against spending the same funds. It is simply a change in the balance assigned to a bank customer's account that prevents the same funds from being spent again.

---

[70] ibid [106].

[71] ibid [109]-[110].

[72] ibid [111]-[113].

[73] It should be noted that this is not necessarily true. The court was probably thinking of the Bitcoin network, where such a practice is used. This is because Bitcoin only records transactions, which is a threat to user privacy. In Ethereum and similar networks, most wallets do not generate new keys for each transaction.

### 3) Capable of assumption by third parties

The third condition concerns respect by third parties for the owner's right to the object in question and the possibility of holding such parties legally liable if they violate this right. Thus, property is a right that affects third parties, not just the parties to a given transaction.[74] In addition, third parties will usually potentially desire such an asset. The fact that the asset has no current market value is irrelevant, for as long as there was a market for it in the past. Thus, there is no doubt that cryptocurrencies can be, and many are, traded on the market.[75]

### 4) Some degree of permanence or stability

The last criterion requires that the object be stable to some degree. This seems to be a consequence of meeting the previous requirements. Undoubtedly, some assets are characterized by low stability, such as tickets. A short useful life is then not a problem. Therefore, similar to funds in a bank account, cryptocurrencies meet the above criterion.[76] Also, technology such as blockchain helps ensure the stability of cryptocurrencies, as the entire history of such assets is verifiable through publicly available distributed registries.[77] According to the court, cryptocurrency remains stable until the private key holder spends it.[78] The risk of theft by unauthorized access to the private key does not affect this. This is because it is not significantly different from the risk of theft of funds in a bank account secured by a password or PIN.[79]

---

[74] David Fox, 'Cryptocurrencies in the Common Law of Property' (25 August 2018) <https://dx.doi.org/10.2139/ssrn.3232501> accessed 19 January 2023.

[75] *Ruscoe* (n 67) [114]-[116].

[76] ibid [117].

[77] The above statement clearly suggests that the court viewed cryptocurrencies through the prism of public networks.

Because cryptocurrencies fulfill the above criteria, the court concluded that they are intangible property, stating, following the authors of the Legal Statement on Cryptoassets and Smart Contracts, that cryptocurrencies are 'a conglomeration of public data, private key, and system rules.'[80] Similar to Sir Simon Bryan, Justice Gendall found that the dichotomous distinction between choses in action and choses in possession does not preclude recognizing cryptocurrencies as property. It would be ironic not to deem as property something with more ownership characteristics than debt.[81] Even if we keep the division into two types of property, cryptocurrencies can be classified as choses in action.[82] According to the court, it is also incorrect to view cryptocurrencies as mere information because their purpose is to create an object of commercial value, not merely to record or transfer knowledge. Despite the lack of a bank guarantee, the combination of a public key and a private key provides a method of transferring such assets, similar to the function of a bank account password.[83] As a result, the court again concluded that, for the reasons stated, cryptocurrencies of all kinds constitute property under Section 2 of the Companies Act 1993 and arguably property in a more general sense.

---

[78] *Ruscoe* (n 67) [118].

[79] ibid [112].

[80] ibid [121].

[81] ibid [124].

[82] ibid [123]-[125].

[83] ibid [122]-[128] cf the comments given earlier with regard to distinguishing the concept of information from data.

However, the Law Commission, as part of its project to reform the law, pointed out that meeting the above conditions of property is not sufficient.[84] It is still necessary that the object of property rights be characterized by certain universal qualities, namely (1) excludability, (2) rivalrousness, (3) separability, and (4) value.[85]

Excludability is the ability to block access to a particular asset.[86] Thus, it is about exercising actual control over a particular asset and deriving certain benefits from it.[87]

Rivalrousness is a characteristic of an object in which one person's use of it negatively affects other people's ability to use it. It boils down to the principle that only one person can hold a certain good.[88] Therefore, information is non-rivalrous. One person's knowledge of a certain fact does not preclude another person from knowing it. However, as I mentioned earlier, information is different from data, while the former is even protected by IPRs.

Separability means that the object must be separated. A man holding a hand in a jacket made to look like a gun cannot be charged with possession of an imitation firearm. One cannot possess something that is not separated from other objects, such as a human hand that is not cut off from the rest of the body.[89]

Finally, value is the most debatable criterion. Of course, it is reasonable to say that the higher the value of an object is, the more protection it deserves. However, the value should not be viewed solely as market value, which is subjective and can fluctuate.[90] I believe that a more appropriate view of the criterion of value (if it has any meaning at all) is that it is a matter of convention that society deems it necessary to protect certain objects. Certain goods should be protected because society expects them to be, not because they are expensive. No one today denies that even sand or gravel can have value, and it is legitimate to protect them, if the other criteria are met, because they can become useful building materials.

---

[84] cf The Law Commission, 'Law Com No 256: Digital Assets: Consultation Paper' (Law Commision 28 July 2022) 22–28.

[85] ibid 21–22.

[86] ibid 29.

[87] Kevin Gray, 'Property in Thin Air' (1991) 50 The Cambridge Law Journal 252, 268.

[88] Joshua AT Fairfield, 'Tokenized: The Law of Non-Fungible Tokens and Unique Digital Property' (2022) 97 Indiana Law Journal 1261, 1263.

[89] *R v Bentham* [2005] UKHL 18 [8].

All of the above criteria seem valuable for resolving the question of the objects of property rights. However, the *Ruscoe v Cryptopia Ltd* case shows that they all fall within the classical definition of property. Thus, it does not seem necessary to refer to these additional criteria in order to consider that crypto-assets, or more broadly, digital data, are objects of such rights.

The case of *Philip Smith v Torque Group Holdings Ltd (In Liquidation)*, heard in the Court of the British Virgin Islands, also provided additional insight into the issue.[91] Torque operated as an online platform for trading crypto-assets but went into liquidation. Torque's liquidators asked the court to help determine the legal status of Torque's accumulated assets. Based on arguments similar to those in the Cryptopia case, the court held that crypto-assets constitute property under the BVI Insolvency Act 2003. The court agreed that, as a general rule, a person who holds a private key is the owner of the crypto-assets that the key controls. However, ownership of such property must be determined on a case-by-case basis, depending on the additional circumstances. The difference is that the liquidators pointed to the existence of two types of wallets offered on the platform. The court held that different conclusions should be reached depending on which wallet contained the disputed goods:

---

[90] The Law Commission (n 84) 33.

[91] *Philip Smith v Torque Group Holdings Ltd (In Liquidation)* BVIHC (COM) 0031 of 2021.

1) For 'User Trading Wallets' – in which various cryptocurrencies were 'mixed' with each other in different wallets, over which Torque had exclusive control, Torque was their likely owner; and[92]

2) For 'User Personal Wallets' – which, in the court's view, were offered as separate hosting services, where the company merely generated a private key but did not have access to or knowledge of the key afterward, the assets assigned to these wallets were not assets of the company.[93]

However, the case lacks further consideration in this regard, for example, of the possibility of recognizing the relationship between the company and its clients as a trust. Therefore, it is impossible to agree with the arguments that having a private key determines the ownership of the funds collected. Stealing a private key can be compared to stealing keys to an apartment. Such an event does not mean loss of ownership of the property. In fact, a private key constitutes an appurtenance because it allows one to dispose of the accumulated funds, just as having the keys to a car allows one to enter and drive the car. However, this should not affect the state of ownership of the car.

Thus, in order to properly assess the realities of the above case, it would be necessary to further analyze whether the crypto-assets were merely made available for use by the users of the 'User Trading Wallets' or whether they were entrusted to the company to store and manage on behalf of the users of such services, the nature of which is similar to a bank's obligation to its client. There seems to be little doubt today that the transfer of funds held in a bank account by an unauthorized person does not affect the actual amount of the bank's obligation.[94] Nevertheless, the protection of token owners trading on exchanges is provided by the later ruling in *D'Aloia v Persons Unknown*, holding that token owners can be considered constructive trustees if there is a fraudulent misappropriation of crypto-assets.[95]

---

[92] ibid [29]-[30].

[93] ibid [31]-[32].

The above positions of the courts seem to form an established case law in the countries influenced by the law of England and Wales, leading to the recognition of crypto-assets as property.[96] In the case of NFTs, the later decision in *Osbourne v Persons Unknown* confirmed this thesis.[97] The Court explicitly recognized that the previous case law also applies to NFTs. The judgment in *Janesh s/o Rajkumar v Unknown Person ('CHEFPIERRE')* also confirmed this.[98]

### 4.1.2. United States – Wyoming and California

In the United States, the US Constitution does not create property interests. The source of the origin and protection of property interests is in specific rules or agreements, such as state law.[99] With only a few exceptions, federal regulations are the sources of exclusive rights, such as copyright and patents. A detailed discussion of each state's legal system is beyond the scope of this article. Therefore, further analysis will be conducted from the perspective of the law of the first state to address the issue of crypto-assets, Wyoming, as well as California, which has become famous for its digital property jurisprudence.

---

[94] A record in a bank account is merely evidence of the existence of a bank's obligation. Otherwise, the removal of the relevant records, for example, due to the actions of a hacker or a bank employee, would mean the termination of the bank's obligation.

[95] *D'Aloia v Persons Unknown* [2022] EWHC 1723 (Ch) [12] et seq.

[96] See *Wang v Darby* [2021] EWHC 3054 (Comm) [55]; *CLM v CLN* [2022] SGHC 46 [39]-[46]; *ByBit Fintech Ltd v Ho Kai Xin* [2023] SGHC 199 [29]-[39]. For perception of cryptocurrencies as property in the context of bitcoin and ether on the grounds of Proceeds (Recovery) Act 2009 (NZ), see also *Commissioner of Police v Rowland* [2019] NZHC 3314.

[97] *Osbourne v Persons Unknown* [2022] EWHC 1021 (Comm) [13] et seq.

[98] *Janesh s/o Rajkumar v Unknown Person ('CHEFPIERRE')* [2022] SGHC 264 [45]-[78].

[99] *Board of Regents v Roth* 408 US 564, 92 S Ct 2701 (1972).

In response to the need to protect crypto-assets, Wyoming state law included Digital Assets provisions on July 2019.[100] The Wyoming Statutes currently include Chapter 29, DIGITAL ASSETS, under Title 34, 'PROPERTY, CONVEYANCES, AND SECURITY TRANSACTIONS.' The regulation defines a digital asset as a representation of economic, proprietary or access rights stored in a computer-readable format.[101] There are three mutually exclusive categories of such assets:

1) Virtual currency – used as a medium of exchange, units of account, or stores of value, that are not recognized as legal tender by the US government;[102]

2) Digital securities – that constitute securities, as defined in WS 17-4-102(a)(xxviii), but exclude digital consumer assets and virtual currency;[103] and

3) Digital consumer assets – used or bought primarily for consumptive, personal, or household purposes and include:

   a) Open blockchain tokens constituting intangible personal property as otherwise provided by law; and

   b) Any other digital assets that do not fall within virtual currency and digital securities.[104]

---

[100] Wyoming Bill No: SF0125, LSO No: 19LSO-0608, Enrolled Act No: SEA No 0039 (2019).

[101] Wyoming Statutes (WS) § 34-29-101(a)(i).

[102] WS § 34-29-101(a)(iv).

[103] WS § 34-29-101(a)(iii).

[104] WS § 34-29-101(a)(ii).

Currently, the most commonly created NFTs fall mainly within the last category. Interestingly, all the above assets have been classified as intangible personal property, subject to specific treatment under Articles 8 and 9 of the Wyoming Uniform Commercial Code.[105] The above regulations mean that in Wyoming, the investor is the direct owner of such an asset without the need for intermediaries.[106] Nonetheless, the regulation has received significant criticism from the Uniform Law Commission because it conflicts with the 2017 proposed regulation for virtual currency businesses, which eventually was not widely adopted.[107]

In 2022, the Commission attempted again to unify crypto-asset regulations, specifically proposing Article 12 of the UCC for 'Controllable Electronic Records' (CER).[108] CER is an electronic record that can be controlled.[109]

The control consists of four conditions: (1) the power to enjoy 'substantially all the benefit,' (2) the exclusive power to prevent others from enjoying those benefits, (3) the exclusive power to transfer control, and (4) the ability to identify as having the above powers.[110] Subject to the exceptions in Section 12-105(c), control for the above purposes is also exclusive when the power is shared with another person (it may also be exercised through another person) and when technical considerations limit the use of the record, for example, by programmed functions that cause the record to be transferred or the resulting benefits to be modified. If a

---

[105] WS § 34-29-102(a).

[106] WS § 34-29-103.

[107] See Matthias Lehmann, 'National Blockchain Laws as a Threat to Capital Markets Integration' (2021) 26 Uniform Law Review 148, 162 et seq.

[108] Uniform Law Commission, 'Uniform Commercial Code Amendments (2022)' (Uniform Law Commission).

[109] ibid, s 12-102(1).

[110] ibid, s 12-105.

person has the powers specified in conditions 2 and 3, the powers are presumed to be exclusive.[111]

CERs include various digital assets, including cryptocurrencies and tokens, regardless of the technology used. Article 12 of the UCC seeks to provide rules for transactions in digital assets. The purchaser of CERs acquires all rights that the transferor had or could have transferred.[112] Bona fide purchasers of CERs receive them free of property claims.[113] The practical application of these rules raises many questions, so we should soon see some interesting rulings in this area and beyond.[114] However, the regulations do not address whether the purchaser of the token owns it and what rights it has.[115]

Thus, the ownership of tokens is much more complicated. Jurisprudence related to recognizing various digital goods as objects of property rights may become the basis for seeking protection. In 2003, the US Court of Appeals, Ninth Circuit, decided a dispute over a domain name and held that the domain name could constitute property subject to the tort of conversion.[116] Kremen sued Network Solutions Inc for illegally transferring the properly registered domain name sex.com to a third party based on a forged letter of its purported re-registration.

---

[111] ibid, s 12-105(b)-(e).

[112] ibid, s 12-104(d).

[113] ibid, s 12-104(e).

[114] Eg *Kleiman v Wright,* Case No 18-cv-80176-BLOOM/Reinhart (SD Fla 2022) (awarding the plaintiff damages for 'conversion of intellectual property' made by one of the alleged co-creators of Bitcoin); cf *Wright v BTC Core* [2023] EWHC 222 (Ch).

[115] Uniform Law Commission (n 108) 231–33.

[116] *Kremen v Cohen* 337 F3d 1024 (9th Cir 2003).

As a condition for benefiting from a claim under the tort of conversion, the claimant must be shown to have had property rights to dispose of the unlawfully used property. Given its positions in prior litigation, Network Solutions stated that the right to use a domain name is intangible personal property.[117] The court agreed with this position, stating, following the *Downing v Municipal Court* decision, that property is a broad concept that includes 'every intangible benefit and prerogative susceptible of possession or disposition.'[118]

To determine whether property rights existed, the three-step test from the case of *G.S. Rasmussen & Associates v Kalitta Flying Ser* was applied:[119]

1) There must be an interest capable of precise definition – The essence of this condition boils down to recognizing a particular good that can be determined with sufficient precision. As in the case of shares in a company's stock or real estate, the right to a domain name is a well-defined interest. The one who registers the domain name obtains the right to determine where a user will go when the user types the domain name into the browser;[120]

2) The object must be capable of being exclusively possessed or controlled – 'The air we breathe, scenic views, the night sky, the theory of relativity, and the friendship of others cannot be reduced to possession and therefore cannot be the basis of property rights.'[121] However, this criterion is met by domain name ownership, since the decision to direct the said Internet traffic is made solely by the registrant. The court

---

[117] ibid 1029; see *Network Solutions Inc v Clue Computing Inc* 946 F Supp 858 (D Colo 1996); *Network Solutions Inc v Umbro International Inc* 259 Va 759, 529 SE2d 80 (Va 2000).

[118] *Downing v Municipal Court* 88 CalApp2d 345, 198 P2d 923 (Cal Ct App 1948).

[119] *GS Rasmussen & Associates v Kalitta Flying Ser* 958 F2d 896 (9th Cir 1992).

[120] *Kremen* (n 116) 1030.

[121] *GS Rasmussen & Associates* (n 119) 903.

also noted that domain names, like other forms of property, are traded. Moreover, they are bought and sold, often for millions of dollars; and[122]

3) The putative owner must have established a legitimate claim to exclusivity – The registration of domain names has been compared to the registration of real estate in the relevant office. This registration ensures that the owner of a particular domain name can be determined. Domain name owners often invest significant resources and time in developing and promoting websites that benefit from operating under a particular domain name.[123] Thus, private property provides an incentive to create and maintain certain goods, which affects the development of the economy and, in this case, the Internet.

Consequently, Kremen was found to have a property right in its domain name and that such a right was an 'intangible property right.'[124] This view of domain names as property has been upheld in cases such as *Office Depot Inc v Zuccarini*,[125] *CRS Recovery Inc v Laxton*,[126] and *Sprinkler Warehouse Inc v Systematic Rain Inc.*[127] There are also similar positions in

---

[122] *Kremen* (n 116) 1030.

[123] ibid.

[124] ibid; see Jay Prendergast, 'KREMEN v. COHEN: THE "KNOTTY" SAGA OF SEX.COM' (2004) 45 Jurimetrics 75.

[125] *Office Depot Inc v Zuccarini* 596 F3d 696, 701–02 (9th Cir 2010).

[126] *CRS Recovery Inc v Laxton* 600 F3d 1138, 1144 (9th Cir 2010).

[127] *Sprinkler Warehouse Inc v Systematic Rain Inc* 859 NW2d 527, 531 (Minn Ct App 2015).

Canadian jurisprudence: in Ontario, *Tucows.com Co v Lojas Renner SA*,[128] and in British Columbia, *Canivate Growing Systems Ltd v Brazier*.[129]

The Cryptopia Ltd case's reasoning clearly confirms that the tokens meet all of the above conditions.[130] This is supported by the ruling in *Shin v Icon Found* that the plaintiff had made a prima facie case that he had an actual and legal interest in pursuing claims for conversion and trespass to chattels due to the freezing of his ICX tokens in the ICON wallet.[131] This case also raised a fundamental issue regarding the original appropriation of tokens. The court correctly concluded that, as a general rule, the original owner should be the person who took the actions necessary to generate certain tokens (for example, mining them according to the rules of the network).[132]

In general, the position affirming the recognition of tokens as property is also supported by the *HashFast Technologies LLC v Lowe* decision, which held that bitcoins could be treated as property in bankruptcy proceedings, such that the trustee can claim the return of its bitcoins under 11 US Code § 550(a).[133] In addition, some cases have held that bitcoins may be subject to forfeiture.[134] It is also worth noting that even for tax purposes, the Internal Revenue Service (IRS) treats cryptocurrencies as property.[135]

---

[128] *TucowsCom Co v Lojas Renner SA* 2011 ONCA 548 [42] et seq. See also *Hanger Holdings v Perlake Corporation SA* [2021] EWHC 81 (Ch) [74].

[129] *Canivate Growing Systems Ltd v Brazier* 2020 BCSC 232.

[130] See J Dax Hansen and Joshua L Boehm, 'Treatment of Bitcoin Under U.S. Property Law' (2017).

[131] *Shin v Icon Found* 553 F Supp 3d 724, 729–36 (ND Cal 2021).

[132] ibid 731.

[133] *HashFast Technologies LLC v Lowe (In re HashFast Technologoes LLC)*, Bankr Case No 14–30725DM, Adv Pro No 15-3011DM (Bankr ND Cal, 19 February 2016).

However, it is essential to determine the proper relationship between the authorized person and a specific asset. As emphasized earlier, having a private key does not determine one's ownership of crypto-assets, just as having keys to an apartment does not necessarily mean that one owns it. However, the courts in *Archer v Coinbase Inc*[136] and *Bdi Capital LLC v Bulbul Investments LLC*[137] took a different view. Both rulings held, as in *Philip Smith v Torque Group Holdings Ltd*, that lack of control over a private key means a lack of ownership of the assets assigned to it. However, the decisions in these cases seem to have been more expressions of the sense of justice not to impose 'a major new duty on all cryptocurrency exchanges (…) to affirmatively honor every single bitcoin fork.'[138]

Thus, exercising control over the token in other ways, such as through a user account on a cryptocurrency exchange, should be possible. It could also be assumed that such a beneficiary should be seen as a tenant or usufructuary of a virtual good. A professional entity such as an exchange would be expected to have an appropriate degree of commitment, just as a bank would be expected to custody valuables. This does not necessarily mean that every fork is automatically respected. However, at the request of the owner of the funds, the exchange should allow the owner to take appropriate action to obtain the resulting alt-coins, for example, by delivering the corresponding private key or transferring the alt-coins to the specified address after the user has paid the necessary costs of handling this process. If the exchange disagrees with this approach, it can sell derivatives instead of crypto-assets.

---

[134] Eg *United States v 50.44 Bitcoins*, Civil Action No ELH-15-3692 (D Md, 31 May 2016); *United States v 89.9270303 Bitcoins,* No SA-18-CV-0998-JKP (WD Tex, 11 February 2022).

[135] IRS, 'IRS Notice 2014-21, IRB 2014-16' (25 March 2014); IRS, 'Frequently Asked Questions on Virtual Currency Transactions | Internal Revenue Service' <https://www.irs.gov/individuals/international-taxpayers/frequently-asked-questions-on-virtual-currency-transactions> accessed 19 January 2023.

[136] *Archer v Coinbase Inc* 53 CalApp5th 266, 267 Cal Rptr 3d 510 (Cal Ct App 2020).

[137] *Bdi Capital LLC v Bulbul Investments LLC* 446 F Supp 3d 1127 (ND Ga 2020).

[138] ibid 140; see Tommy Goodwin, '"Not Your Keys Not Your Coins" Becomes the Law of the Land in California.' (*Medium*, 14 April 2021) <https://medium.com/coinmonks/not-your-keys-not-your-coins-becomes-the-law-of-the-land-in-california-689626b06bfc> accessed 19 January 2023.

### 4.2. Civil law, including Poland and Germany

Unlike common law systems, legal systems derived from Roman law approach the problem of property differently.[139] German law and legal systems based on it (including Polish law) distinguish between tangible and intangible property. As a rule, ownership of things is understood only as authority over tangible objects.[140] But this is not exclusive. Austrian law, for example, treats anything distinguishable from and useful to a person as a thing.[141] Some authors suggested that Italian law, which defines goods as things that can be subject to rights, and French law, which purports to regulate tokens as digital assets, may also provide some solutions.[142] The perception of digital assets as property protected by private law was also recognized in 2019 by a Chinese court.[143] The Shanghai High People's Court also published a document stating that bitcoin constitutes a virtual object of property rights.[144]

---

[139] Rahmatian (n 9) 187.

[140] '*Sachen im Sinne des Gesetzes sind nur körperliche Gegenstände.*' (Only corporeal objects are things as defined by law), Bürgerliches Gesetzbuch (BGB), para 90; '*Rzeczami w rozumieniu niniejszego kodeksu są tylko przedmioty materialne.*' (Things within the meaning of this Code are only tangible objects), Ustawa - Kodeks cywilny 1964 (Polish Civil Code), art 45.

[141] '*Alles, was von der Person unterschieden ist, und zum Gebrauche der Menschen dient, wird im rechtlichen Sinne eine Sache genannt.*' (Everything that is distinct from the person and serves for the use of people is called a thing in the legal sense,) Allgemeines Bürgerliches Gesetzbuch (ABGB), para 285.

[142] Rosa M Garcia-Teruel and Héctor Simón-Moreno, 'The Digital Tokenization of Property Rights. A Comparative Perspective' (2021) 41 Computer Law & Security Review 1, 5.

In 1937, Polish law was one step away from forging the concept of property similar to the common law system. That year, the Codification Commission presented the Draft of the Property Law.[145] According to Article 17, rights *in rem* also applied to goods that are not things, such as works, enterprises, claims, or other legitimate interests, and even electricity or other forms of energy. However, this chapter did not contain a comprehensive regulation. Only two other provisions were included. Article 18, which stipulated that for such assets, appurtenances may be personal property and other goods that are not things; and in the absence of specific provisions, the provisions on things shall apply to them. Furthermore, Article 19 stipulated that for the fruits of such goods, the provisions on the fruits of things shall apply mutatis mutandis unless there are special rules in this regard.

---

[143] *Li and Bu v Yan, Li, Cen and Sun*, (2019) Hu 01 Min Zhong No 13689/(2019) Hu 0112 Min Chu No 12592 ((2019)沪01民终13689号/(2019)沪0112民初12592号); see Shanghai Minhang Court's statement on the protection of virtual property (virtual property) of 17 August 2021 "典"亮闵法的, '如何保护看得见摸不着的虚拟财产？—"典"就会' (微信公众平台) <https://mp.weixin.qq.com/s/Jb-mG7rJ0PN3hU4v8Y9XFg> accessed 19 January 2023.

[144] As reported on the Sina website '上海高院：比特币作为虚拟财产具有财产属性，受财产权法律规范的调整_新浪财经_新浪网' (5 May 2022) <https://finance.sina.com.cn/money/lczx/2022-05-05/doc-imcwiwst5740140.shtml> accessed 19 January 2023; see '比特币是否具有财产属性？怎样执行返还交付？' (5 May 2022) <https://finance.sina.com.cn/money/bank/bank_hydt/2022-05-05/doc-imcwipii8115103.shtml> accessed 19 January 2023; see also Civil Code of the People's Republic of China (Minfadian), art 127 (English translation available at <http://english.www.gov.cn/atts/stream/files/5feda5b8c6d0cc300eea77ac> accessed 19 January 2023.

[145] Polska Komisja Kodyfikacyjna Podkomisja Prawa Rzeczowego, *Komisja Kodyfikacyjna. Podkomisja Prawa Rzeczowego. Z. 1* (Wydawnictwo Urzędowe Komisji Kodyfikacyjnej 1937).

Fryderyk Zoll Jr, who mainly initiated the above-mentioned bill, explained the goals of introducing such a regulation.[146] Zoll stated that Roman law already counted *res incorporales* as property, which included inheritance (*hereditas*), usufruct (*ususfructus*), and obligations (*obligatio*).[147] Many years later, such reasoning could be found in many laws that were once in force in the Polish lands. As an example, the author mentions the Prussian Landrecht, which considered an object of property anything that someone could exclusively use; the French Code, which refers to goods (*biens*) instead of things; and the aforementioned Austrian Civil Code. However, Zoll deemed the above-mentioned laws as incomplete and as treating the problem only incidentally.[148] First, it is unreasonable to say that estate consists of things and rights, because only property rights can be part of estate.[149] Therefore, the draft property law stipulated that estate division should be based on differences in the substance of property rights rather than on the objects of those rights and divided the estate into two primary groups: (1) rights *in rem* and (2) claims. Rights *in rem* are supposed to be rights under which an object is submitted to the direct authority of a particular person, effective against others.[150] This object was supposed to be any good that represents an assessable value in money and can be subjected in whole or in part to the exclusive authority of a private person.[151] Such an approach is similar to the common law concept of property. On the other hand, the drafters understood claims as the right to demand a certain performance from a particular person.[152]

---

[146] Fryderyk Zoll Jr, 'Przedmiot Praw Rzeczowych' (1938) 1 Kwartalnik Prawa Prywatnego 3.

[147] ibid 3–5.

[148] ibid 5–8.

[149] ibid 11.

[150] ibid 11, 14–18.

[151] ibid 12.

However, the above rules did not enter Polish legislation, which defines property narrowly. Some authors postulate that this does not prevent the application of property law to intangibles by analogy.[153] While such a solution is needed, it is impossible to ignore the legal definition of things, which, as in German law, uses the wording that these provisions apply 'only' to tangible objects. Meanwhile, crypto-assets fall into the broader category of virtual goods (i.e. intangible objects and currencies in online communities).[154] According to the general opinion of the scholars, such goods are not things under Article 45 of the Civil Code, which is also confirmed by Polish case law on domain names or PINs.[155]

To a certain extent, Polish law recognizes the possibility of new property rights. Article 44 of the Civil Code states that ownership and other economic rights should be included in the broader estate concept. In tax cases, such ambiguous wording has served the courts to classify cryptocurrencies as an economic right that can be sold but does not constitute a property right.[156] Thus, the exchange of one cryptocurrency for another was considered a contract for the exchange of 'intangible economic rights.'[157] However, it is impossible to determine what kind of economic rights these should be. This issue is also problematic in other countries. For example, the Supreme Court in Spain has ruled that bitcoins cannot be returned because they are merely intangible units of account in a distributed network of the same name. However, the court can order compensation in fiat currency.[158]

---

[152] ibid 11.

[153] Kacper Górniak, 'Prawo Własności Jednostek Waluty Kryptograficznej' (2019) 28 Kwartalnik Prawa Prywatnego 561, 604–07.

[154] See Jason Fernando, 'Virtual Good' (*Investopedia*, 21 July 2022) <https://www.investopedia.com/terms/v/virtual-good.asp> accessed 19 January 2023; Jakub Wyczik, *Dobra wirtualne z perspektywy użytkowników i dostawców treści. Prawne uwarunkowania obrotu* (CHBeck 2022) 42.

[155] Wyczik (n 154) 57–64.

[156] Najwyższy Sąd Administracyjny [Polish Supreme Administrative Court], 6 March 2018, II FSK 488/16.

[157] Najwyższy Sąd Administracyjny [Polish Supreme Administrative Court], 27 April 2022, II FSK 2181/19.

From the view of the general systematics of private law, since we cannot speak of property, perhaps we should refer to claims. However, the sheer determination of the content of a claim is difficult. It would have to consist of a specific obligation of some entity to the buyer to perceive them as the owner of a given good. It is impossible to establish the source of such an obligation other than contract law, which may prove unreliable when interpreting most transactions.[159] In addition, it would be necessary to create at least the obligation to:

1) store the token in a smart contract;

2) ensure that it can be disposed of; and

3) enable the use of its functions (for example, when the token allows one to use it to produce a new token).[160]

Moreover, in adapting the concept of obligations to tokens, the fundamental problem is determining who the debtor is. Given the nature of crypto-assets, there is typically no central administrator. Thus, it appears that no entity is obligated to a potential creditor. Assuming a model state in which there are no intermediaries (for example, exchanges), a potential debtor can be sought among the following three entities:

1) all nodes of the network;

---

[158] Tribunal Supremo [Spanish Supreme Court] STS 2109/2019 ECLI:ES:TS:2019:2109.

[159] They usually occur without additional terms and conditions or use ambiguous phrases that suggest the purchase of certificates, titles, tags, or digital content. See Juliet M Moringiello and Christopher K Odinet, 'The Property Law of Tokens' (2022) 74 Florida Law Review 607, 632 et seq.

[160] Particularly in online games, tokens can represent virtual resources for crafting items.

2) the 'owner' of the smart contract in which the tokens are stored; and

3) the issuer of a specific token.

The first concept requires recognizing that all persons who join the network by becoming its nodes make a kind of cumulative accession to the debts of the other nodes. However, this accession to the debt must be made at least under a condition that the debtor can withdraw at any time by ceasing to act as a node. Alternatively, all creditors would have to consent to the debtor's unilateral decision to discharge the debt at any time. This contradicts the economic modus operandi of crypto-assets and would also allow for a release from liability in case of a dispute. Moreover, while such a debtor may seem somewhat appropriate for cryptocurrencies, it does not match the nature of tokens. Network nodes are to smart contracts what hosting services are to websites. They merely provide computing power, so authorized individuals can invoke certain functions in such programs.[161] Consequently, they cannot be considered debtors to the token owners, and at most, they could have some contractual relationship with the smart contract owner.

Therefore, the above concept leads one to consider the debtor as the smart contract owner. As a rule, every smart contract with tokens has an owner, so it can be assumed that such an owner should be the guarantor of the content (tokens) of the smart contract, just like the website operator. However, this is not a simple solution because, technically, changing the address of the smart contract owner involves calling a single command that assigns such a role to the new address. Formally, this should be considered a debt assignment to a new person, but in practice, such a change does not require the consent of the new owner. The change is purely technical, so there is nothing to prevent the smart contract from being abandoned and assigned to an inaccessible address. But the debt cannot be unilaterally released, and even if it could be, the users would not benefit from such a relationship. Therefore, while such a solution is more appropriate for tokens, it still does not provide satisfactory results.

---

[161] The Ethereum network uses structures called 'Merkle-Patricia-Tries' for data storage. The overall state of the network is recorded within the World State Trie, while smart contract data are stored within the Storage Trie. See Kamil Jezek, 'Ethereum Data Structures' [2021] arXiv:210805513 [cs].

The third option, that the debtor is the issuer of the token, assumes that the token is a claim of its holder to its issuer. This is analogous to negotiable instruments or documents of title. However, this solution does not seem to be valid. As in the case of securities, the issuer acts as a guarantor of the enforceability of the claims that the securities represent. Thus, the token would always have to be such an instrument, which is not necessarily true. Moreover, a token as an object of trade in such a relationship is the equivalent of a document rather than the obligation of its issuer. Thus, it cannot be said that the document issuer is responsible not only for their debt but also for ensuring that the document holder does not lose it or have it stolen. As with the second option, if the issuer ceases to exist (for example, liquidated), this does not mean that the tokens are automatically destroyed. They will probably lose their function but can still serve as a collector's item.

Regardless of which entity is considered the debtor, the claim concept does not correspond to the nature of the right to dispose of virtual goods. The fact that a claim is a relative right, effective only against certain persons, is a significant disadvantage. If the debtor ceases, the token should stop being traded. Perhaps the debtor's liquidators could terminate trading in the token even earlier. Similarly, if the token is stolen, the original owner would lose control of the record in the registry, but should still be considered a creditor. Therefore, such a creditor could not file a claim against the thief because the change in the database would not affect the existence of the claim. The thief could only be sued for potentially interfering with other claims resulting from holding a token (to obtain further benefits, if applicable).

At the same time, as mentioned above, this analysis is based on a theoretical model without intermediaries. However, as a rule, there are exchanges, issuers, owners of smart contracts, providers of token storage services, token acceptors, and other persons obliged to provide certain services. Perhaps for most relationships, a token should be seen as a claim to a claim to a claim, and so on. This makes the situation so complicated that it is impossible to clearly state which claim is the subject of the transaction and whether the transfer of one claim entails the transfer of the others. Therefore, creating such complex relationships is not beneficial to the security and certainty of legal transactions.

Another important problem is determining the debtor's identity to enforce claims against them in court. Publicly available information does not make it possible to identify the parties to the transaction by name. A procedural solution may be to serve the claim using a token sent to the alleged infringer's address.[162] However, in many countries, local regulations would not allow such a delivery.

In conclusion, the claim construction is entirely unsuitable for solving the problem of token ownership. By its very nature, a claim does not correspond to an exclusive right to control a given good. This position has already been recognized in the context of the issue of virtual goods.[163] A new legal framework must be created for such goods.[164]

### 4.3. Lex rei sitae of tokens

Based on the above considerations, the determination of the law applicable to the above categories of rights and infringements remains essential. This issue has been the subject of many of the aforementioned disputes. However, due to the limited scope of this article, I will focus on the concept of *lex rei sitae* of crypto-assets using one of the rulings in England as an example. It is also worth noting that a highly intriguing but controversial proposal has been made by the UNIDROIT Working Group on Digital Assets. The proposal assumes that the law applicable to proprietary issues should first be decided by the law of the country expressly specified in the digital asset or, in the absence thereof, by the law expressly indicated in the system or platform on which the digital asset is recorded.[165] By acquiring, transferring, or otherwise dealing with a digital asset, a person is deemed to have agreed to the law specified in this manner. Similarly, the 2022 UCC Amendments follow the existing choice of law rules for financial assets. The applicable law is generally understood to cascade from the jurisdiction expressly specified at the time of purchase by the CERs or other associated or attached records, through the jurisdiction of the system in which the CER is recorded, and otherwise as the jurisdiction of the District of Columbia.[166] While the idea of the choice of law seems logical, such a solution is not necessarily promising. Tokens are closer in their location to IPRs than to typical objects of property rights, as explained below.

---

[162] See *D'Aloia* (n 95) [38]-[40].

[163] For legal characterization of claims in the case of virtual goods and their sources and types, see Wyczik (n 154); Fairfield (n 88) 1261–313.

[164] Wyczik (n 154) 183–88.

The July 2021 verdict in *Fetch.AI Ltd v Persons Unknown Category A*[167] concerned, in particular, the determination of the law applicable to the breach of confidence in a private key. The basis for determining the applicable law became the provisions of the Rome II

---

[165] In doing so, consideration should be given to records attached to or associated with the digital asset, or the system or platform, if they are readily available to those dealing with the relevant digital assets, see UNIDROIT Digital Assets and Private Law Working Group, 'Study LXXXII – Draft UNIDROIT Principles on Digital Assets and Private Law' (UNIDROIT January 2023) 21–25.

[166] Uniform Law Commission (n 108) 260–65.

[167] *FetchAI Ltd v Persons Unknown Category A* [2021] EWHC 2254 (Comm).

Regulation.[168] Under the Brexit Act,[169] it remains applicable in the UK to the extent that it has not been subsequently amended.[170] It was held that a claim for the breach of confidence falls within the scope of Article 4(1) of Rome II and does not constitute a case of violation of privacy or personality rights, which are excluded from the Regulation. In doing so, the application of Article 6 Rome II was also explicitly rejected, as the case did not involve an act of unfair competition.[171]

Therefore, the court had to determine the law of the country where the damage occurred. The court found that this required finding the location of the crypto-assets. For this purpose, the court referred to the decision in *Ion Science v Person Unknown*, in which paragraph 13 held, citing Andrew Dickinson, that the *lex situs* is that of the place where the owner resides.[172] Thus, English law applied, since the first applicant was domiciled in England.[173] However, such a simplistic view of crypto-asset location raises some doubts. The owner's place of residence alone does not seem decisive. Determining the applicable law under Rome II based on domicile contradicts the rules for determining the place of damage developed in the case law of the Court of Justice of the European Union (CJEU).[174] Meanwhile, according to Section 6(3) of the Brexit

---

[168] Regulation (EC) No 864/2007 of the European Parliament and of the Council of 11 July 2007 on the law applicable to non-contractual obligations (Rome II) [2007] OJ L311/40.

[169] European Union (Withdrawal) Act 2018.

[170] See The Law Applicable to Contractual Obligations and Non-Contractual Obligations (Amendment etc) (EU Exit) Regulations 2019.

[171] *FetchAI* (n 167) [11]-[13].

[172] ibid [14]; Andew Dickinson, 'Cryptocurrencies and the Conflict of Laws' in David Fox and Sarah Green (eds), *Cryptocurrencies in Public and Private Law* (Oxford University Press 2019) para 5.109.

[173] Regarding the second plaintiff, which is a company registered and operating in Singapore, the court concluded that the first plaintiff is acting as an agent of that company, which is sufficient to apply English law.

[174] Case C-304/17 *Helga Löber v Barclays Bank plc* [2018] ECLI:EU:C:2018:701, paras 16-36.

Act, the decisions of the CJEU constitute 'retained case law.' However, even if the court's decisions are not entirely precise, the outcome is still consistent with the case law of the CJEU.

The fact that the case involved the assets of a company based in England is not enough to establish jurisdiction. If the infringer can act practically anywhere in the world, it is impossible to consider that the place where the damage is caused is only the country of origin of the injured party or the country from which the infringer acted.[175] A token is a certain social construct, similar to IPRs. Therefore, the infringer should expect that when acting over an electronic network, as in the case of IPRs, the damage can be suffered by virtually any entity in the world that was a legitimate owner of such an asset in a given jurisdiction. This is because the damage is not to the property, but to the right granted to the owner of that property. This right is always intangible and has no location. Therefore, the applicable law for the protection of such property is its *lex rei sitae*.

Since the plaintiff claimed a trespass to their goods, which were protected by English property law, the damage arose in England. This is because the damage affected the ability to dispose of the goods under English law. Therefore, protection in this regard should take a form similar to the *lex loci protectionis*, which is justified by the impossibility of determining the location of the intangible good. The criterion of *lex rei sitae,* traditionally understood as applying to tangible things, loses its meaning in the case of virtual goods.[176] It is therefore necessary to develop another universal standard, which may be the concept of a specific *lex causae* presented above, referring to the principle of *lex loci protectionis*.

---

[175] See ibid para 35; Case C-375/13 *Harald Kolassa v Barclays Bank plc* [2015] ECLI:EU:C:2015:37, para 56; Case C-168/02 *Rudolf Kronhofer v Marianne Maier* [2004] ECLI:EU:C:2004:364, para 20.

[176] See Amy Held, 'Does Situs Actually Matter When Ownership to Bitcoin is in Dispute?' (2021) 4 JIBFL 269.

## 5. Rights to assets that can be linked to the token

Tokens represent an unlimited pool of potential goods and corresponding rights. Thus, certain rights can be considered to accompany tokens as inherent parts of them. These may include the right to benefit from certain performances (for example, receiving goods or services). However, tokens can also be created to represent rights such as ownership of things, shareholder rights, or copyright.

Tokens can also represent rights to other digital data. A token is simply a specific value or unique identifier associated with a user's account address. As in the case of online games, users most often identify such a database entry with the ability to dispose of a specific virtual good. Technically, it is incorrect to equate such entries with the digital assets they potentially represent, such as 3D models. Virtual goods can be independent objects of property rights.[177] Thus, rights to a virtual good should establish the existence of virtual property objects. In such a case, we can discuss several related assets that can be independent trading objects.

A token can also be a key to control another property. Thus, it can be the equivalent of a movable object, such as a key to an apartment. Similarly, a token can provide access to other resources by acting as a credential for digital files, for example. Note, however, that hypothetically nothing prevents the owner of the key from being someone other than the owner of the apartment. Thus, the transfer of token may be merely a means by which parties to a contract express their will to transfer rights, just as the transfer of a ticket by a movie theater employee in exchange for payment results in a contract to see a movie. Therefore, each time tokens are traded, the relationship between the assets involved must be examined on a case-by-case basis.

---

[177] See Joshua Fairfield, 'Virtual Property' (2005) 85 Boston University Law Review 1047.

It is also important to remember that many tokens are not linked to any other assets. Sometimes, a particular value is derived from a token as a standalone asset. Some authors refer to such tokens as 'endogenous' and those linked to some assets as 'exogenous.'[178] Therefore, different types of such links must be distinguished, and general conclusions should not be drawn without considering individual circumstances.

It should be clear that some rules already allow selected assets to be managed by tethering them to specific records. However, such regulations do not answer the general question of how different assets could be bound to tokens. Some laws simply allow specific classic instruments to exist in a new form. However, any link between several assets must be based on specific regulations. This is also the case with Article 7(1) of the Liechtenstein Token Act, which states that the disposal of tokens results in the disposal of the rights that they represent, which, according to Article 7(3), is generally effective vis-à-vis third parties.[179]

Speaking of linked rights, one will often speak of claims arising from owning a given token. Such obligations are characteristic of the so-called 'utility tokens,' which seem to carry claims to entitlement to some benefit. Such tokens can therefore appear as vouchers or even negotiable instruments.[180]

If we want to consider crypto-assets as incorporating further rights into themselves, we should consider them negotiable instruments or documents of title. As a general category of assets, tokens do not correspond, in the vast majority of cases, to the currently known instruments. Even if they are similar to some instruments, they are created informally, so they do not meet the conditions for being considered, for example, shares or bonds. Therefore, when tokens aim to incorporate, for example, only profit participation right, the basis for the payment may be an external relationship other than share ownership. It could be a promise to anyone who meets certain conditions (for example, becomes a token holder or provides proof of its possession in a certain way) or a contract in which the form of 'payment' will be a particular behavior of the token holder (for example, the token's transfer, delivery, or destruction). Usually, nothing prevents a separate contract from being the source of the token user's claim to receive certain benefits.

---

[178] Michael G Bridge and others, *The Law of Personal Property* (Third, Thomson Reuters/Sweet & Maxwell 2022) para 8-043; Hin Liu, 'Digital Assets: The Mystery of the "Link"' (2022) 3 JIBFL 161.

[179] Token-und VT-Dienstleister-Gesetz (TVTG) 2019, s 7(1), 7(3).

[180] Recognizing that rights to a token also constitute such an instrument would mean creating instruments that include other instruments. Again, the need to distinguish between the right and the object that the right affects should be clearly emphasized.

Sometimes, however, the token can be seen as a simple voucher or similarly functioning negotiable instrument.[181] Note that this does not prevent a given token from being treated as a negotiable instrument by one debtor and as a voucher by another. Nothing prevents another entity from recognizing someone else's document as the basis for their performance. The question, however, is whether we would still be dealing with a voucher or a negotiable instrument within the meaning of the law when we are talking about the perspective of a person who is not its issuer. In such a case, the proper distinction between the respective assets is sufficient. It is difficult to find practical objections to a token's having the effect of a negotiable instrument for one claim and, simultaneously, for another claim, to perform the function of a voucher. After all, in the same way, a bill of exchange or a check in the traditional form is simply a piece of paper, so it can also serve as a document carrying additional information.

---

[181] The mere distinction of these categories is undoubtedly difficult with the concept of an open catalog of such instruments in many civil law countries. I do not see a clear answer as to how a voucher or gift card should be classified so that the criteria for distinction remain objective and reproducible. Perhaps this means that it is time to base such relationships solely on the underlying agreement.

Notwithstanding the above, effective trading of tokens as instruments is obstructed primarily by the archaic regulation of most legal systems, including the Polish Civil Code. As a rule, the owner is not determined by the content of a token itself, but only as a result of mapping addresses with specific records in a database, in the same way in which the owner of a bank, email, or social network account is determined. According to Article 517 § 2 of the Civil Code, a claim from a bearer paper is conveyed by transferring ownership of the bearer paper. In addition, for effective transfer of ownership of such an instrument, the regulations also require handing it over. Therefore, with the current wording of Article 45 of the Civil Code, since an intangible document is not a thing, it is impossible to issue such instruments in a dematerialized form without a special norm. Therefore, a document in digital form could not be the subject of property rights. This means that if a token placed on a distributed public network were to be considered a negotiable instrument, it would be impossible to further transfer economic rights to which the token is linked. The problem with adapting regulations to dematerialized instruments has been known for a long time. Nevertheless, general solutions are still lacking, and for now, this problem is addressed individually for specifically named instruments. It is worth noting that the Electronic Securities Act, which has been in effect in Germany since June 2021, creates a new framework for the operation of electronic instruments with a focus on bearer bonds.[182] An interesting solution is provided by § 2(3) of this act, according to which electronic instruments are considered things under § 90 BGB. This solution seems intuitive and reasonable from the viewpoint of potential investors. However, this applies only to instruments directly regulated by the provisions of the above law.

Separately, tokens can sometimes be considered instruments of entitlement. One might conclude that there is a similar problem with such instruments as there is with negotiable instruments. However, concerning instruments of entitlement, the rule of Article 517 § 2 of the Civil Code applies only mutatis mutandis. Even more importantly, for tokens that act as instruments of entitlement, further rights are not automatically transferred solely as a result of the transfer of ownership and delivery of the document but at most, as a result of the transfer of rights to the tokens themselves.[183] As an instrument of entitlement, a token is merely an evidentiary instrument that allows its holder to prove entitlement to certain benefits from other persons. Therefore, the object of trading to obtain certain benefits can only be a claim or other economic right underlying such an instrument.[184] At the same time, it should be emphasized that even if a document is perceived as a way to use other goods, it is not yet an instrument. In practice, there are many forms of digital data that no one claims to be negotiable instruments, for example, a record in the domain name registrar's database confirms the right to control a domain name, and the ID of a fan page administrator on a social network confirms the right to manage a fan page.

---

[182] Gesetz über elektronische Wertpapiere (eWpG) 2021.

The fact that a token is not an instrument under private law does not mean it is not a security under public law. For the member states of the European Union (EU), financial market supervision is a largely harmonized area, with national regulations mainly aimed at incorporating EU law. This issue is particularly related to Directive 2014/65/EU, which uses the concept of transferable securities, as defined in Article 4(1)(44).[185] According to its wording, such instruments are 'those classes of securities which are negotiable on the capital market, with the exception of instruments of payment (…).'[186] Undoubtedly, the basis of the uniform interpretation of EU law is the principle of autonomy of the concepts used in the legislation.[187] Thus far, the concept of security in this directive has not been considered in detail by the CJEU. Particularly important, however, is the criterion of being negotiable on the capital market, which could effectively limit the recognition of many tokens as securities. Ambiguities in this area are expected to be resolved by the Markets in Crypto-Assets (MiCA) Regulation, which will define a newly regulated market for crypto-assets.[188]

---

[183] See Sąd Najwyższy [Polish Supreme Court], 16 April 2003, I CKN 202/01.

[184] As with coupons, payment cards, or tickets. For Polish law, see Maciej Giaro, 'Transakcje Tzw. Sprzedaży Kuponów w Ramach Zakupów Grupowych' (2014) 3 PPH 40; Michał Kozik, 'Charakter Prawny Kart Płatniczych' (2006) 6 PPH 35; Konrad Zacharzewski, 'Charakter Prawny Dokumentu Uprawniającego Do Przejazdu Ulgowego. Glosa Do Wyroku TK z Dnia 29 Czerwca 2004 r., P 20/02' (2005) 12 PiP 112.

Unlike European regulation, the US system has explicitly recognized the possibility of qualifying certain tokens as securities falling within the conceptual category of an investment contract. Verification is done using the so-called 'Howey test,' which requires four criteria: (i) investment of money, (ii) in a common enterprise, (iii) with the reasonable expectation of profit, and (iv) to be derived from the efforts of others.[189] Not all tokens meet these conditions, but US

---

[185] Directive 2014/65/EU of the European Parliament and of the Council of 15 May 2014 on markets in financial instruments and amending Directive 2002/92/EC and Directive 2011/61/EU (recast) (Text with EEA relevance) [2014] OJ L173/349.

[186] Further listing examples of these securities, such as stocks or bonds.

[187] See Case C-467/08 *Padawan SL v Sociedad General de Autores y Editores de España (SGAE)* [2010] ECLI:EU:C:2010:620.

[188] Regulation (EU) 2023/1114 of the European Parliament and of the Council of 31 May 2023 on markets in crypto-assets, and amending Regulations (EU) No 1093/2010 and (EU) No 1095/2010 and Directives 2013/36/EU and (EU) 2019/1937 (Text with EEA relevance) [2023] OJ L150/40.

[189] *SEC v Howey Co* 328 US 293, 66 S Ct 1100 (1946).

Securities and Exchange Commission (SEC) recognizes DAO tokens as securities.[190] On the other hand, NFTs that represent only certain digital content will not represent an investment of money in a common enterprise. There may be some doubt if the token's creator receives some benefit from further resale. However, considering such an agreement as always satisfying the Howey test would require that all copyright agreements specifying remuneration in royalties be considered securities. The mere earning of a profit from a certain activity cannot be sufficient yet.[191]

The above activities and transactions under private law should be treated separately from the rights to the tokens themselves. In the same way, in a contract of carriage based on a ticket, the carrier undertakes to fulfill its obligation in exchange for the validation of the ticket, which is equivalent to the payment of a price. At the same time, the mere purchase of a ticket does not mean that a contract of carriage is simultaneously concluded between the parties. The ticket appears as an independent good, which the carrier has undertaken to accept as a means of exchange between the parties. Nothing prevents tokens from being viewed in the same way.

American scholars seem to draw similar conclusions. According to Juliet M. Moringiello and Christopher K. Odinet, the mere fact that a token is associated with a picture of another asset (such as a physical sculpture), which it is supposed to represent, does not mean that the purchaser of the token becomes the owner of the represented asset. If a seller transfers the rights to a token to Buyer A but does not yet deliver the sculpture and then later sells and delivers the sculpture to Buyer B, Buyer B will be considered the sculpture's owner.[192] The mere delivery of a token does not result in the transfer of rights to the represented asset, which is done through a separate contract. This is because common law systems recognize property rights as deriving from a superior title to possession, which means that obtaining possession is a condition for transferring rights to personal property.[193] Of course, this does not prevent claims against the original seller for breach of contract. Therefore, the authors emphasized that NFTs do not bind any further rights.[194] The fact that the law imposes further legal effects depending on the possession of certain forms of certificates is due to regulations and not because the contract so provides. Currently, no universally applicable regulations have such effects on tokens.[195]

---

[190] Securities and Exchange Commission, 'Report of Investigation Pursuant to Section 21(a) of the Securities Exchange Act of 1934: The DAO' (Securities and Exchange Commission 25 July 2017). For more on crypto-asset securities, see Lewis Cohen and others, 'The Ineluctable Modality of Securities Law: Why Fungible Crypto Assets Are Not Securities' (13 December 2022) <https://dx.doi.org/10.2139/ssrn.4282385> accessed 23 May 2023.

[191] However, see *Friel v Dapper Labs Inc* 21 Civ 5837 (VM) (SDNY Feb 22, 2023) (finding that NBA Top Shot Moments NFTs may constitute investment contracts, but emphasizing the role of the defendant's private blockchain).

The recently proposed amendments to the UCC are similar. Article 12 applies to records, not to additional rights attached to them. Other laws are supposed to determine what rights are linked to the token (for example, UCC Article 9).[196] Thus, copyright law determines what, if any, rights to the linked artwork are granted to the token purchaser. The exception is the right to payment ('controllable account' and 'controllable payment intangibles') to the person who controls the CER, which the purchaser acquires free of encumbrances.[197] The debtor under such a title may, in principle, discharge its obligation by paying the person who controls CER or the person who formerly controlled CER unless it receives adequate notice from the current controller objecting to payment to a former holder.[198]

---

[192] Moringiello and Odinet (n 159) 658.

[193] ibid 659; *Lanfear v Sumner* 17 Mass 110, 113 (1821). For UK, see Sale of Goods Act 1979, s 17, 24.

[194] Moringiello and Odinet (n 159) 641–43, 660–63.

[195] ibid 642–43.

[196] Uniform Law Commission (n 108) 223–34.

[197] ibid, s 12-104(f), 9-102(a).

As in the U.S., in England and Wales, in principle, any assets accompanying tokens will result only from an additional contract. It is possible to recognize, however, that some tokens will constitute documentary intangibles, which embody the right to demand the performance of obligations recorded in a document. An example is a bill of exchange, which signifies a written agreement by one person to pay a specified sum of money to or upon the order of a specified person. However, the full recognition of tokens as such documents was prevented primarily by the problem of inadequate legal regulation.[199] But Electronic Trade Documents Act 2023 states that digital trade documents are equal to paper trade documents and therefore can be tokenized.

It should also be emphasized that regardless of the applicable law, in the absence of the existence of a link between the assets, there are, in principle, no grounds for prohibiting the purchaser of a token that accompanied the transaction of acquiring another good from continuing to dispose of the good independently.[200] A particular way to promote certainty in trading may be the introduction of a legal presumption that the person who exercises actual control over the token also has title to the goods that the token represents.

Juliet M. Moringiello and Christopher K. Odinet also pointed out that most contractual terms contradict the existence of the aforementioned bond between the token and another asset.[201] However, such a general simplification can lead to imprecise distinctions between specific transactions, as shown by the not-so-clear attempt to seek answers in the terms and conditions of trading platforms.[202] After all, the providers of such platforms are not parties to the transaction, but the people who offer to sell specific tokens are. Information on whether a token is linked to other assets should be mainly included in the terms and conditions of sale attached to specific tokens. However, most such transactions do not include additional contractual provisions.

---

[198] ibid, s 12-106.

[199] The Law Commission (n 84) 304–09. However, for proposal to equate some electronic forms of such instruments with paper, see The Law Commission, 'Law Com No 405: Electronic Trade Documents: Report and Bill' (Law Commision 15 March 2022).

[200] Bridge and others (n 178) para 5-026. Of course, it is possible to contractually prohibit such actions, but the breach of this provision will not, by itself, affect title to a particular asset.

Therefore, the only reasonable claim is to consider that additional benefits come from separate contracts, which may or may not accompany the contract for the token sale. However, the parties' behavior should be interpreted with caution, since the value of most tokens is derived from their accompanying assets. It would be reasonable to make a presumption that the parties also intend to transfer the rights to the normal use of the accompanying assets for their intended purpose. This does not mean, however, that by purchasing a token that represents a photograph of a portrait of the Mona Lisa, the purchaser expects to transfer the original photograph, the copyright to the work, and the original painting. Instead, such an act is seen rather as buying a digital copy of the work for saving as a file on one's device or displaying on a screen. However, this raises the issue of the need to establish ownership of digital copies of works.

---

[201] Moringiello and Odinet (n 159) 643.

[202] ibid 628 et seq.

**6. Intellectual property rights embedded in or otherwise associated with the token**

When discussing token rights, there is an additional need to distinguish IP that is merely embedded or associated with a token differently than in layer two. Depending on individual circumstances, the transfer of a token may also result in the transfer of such rights.

Copyright and related rights are undoubtedly independently protected assets. The mere sale of a copy of a book does not automatically transfer the copyright.[203] Along with the book sale, however, there may be a parallel copyright transfer. Therefore, the terms of each transaction must be analyzed case by case.

The need for such a transfer can be found in various Web3 projects. An example is the marketplace of Decentraland, which attempts to impose on its users the obligation to transfer all their rights to the buyer. This is because, as Decentraland's terms and conditions state that NFT sales transactions will covey the title, ownership, and IPRs over NFT to the buyer, and the creator waives moral rights to the fullest extent possible.[204] Similarly, some smart contracts may have copyright transfer metadata, such as those generated using Mintable, which allows the use of the copyright_transfer boolean variable.[205] The question is whether such solutions will effectively transfer copyright.

First, naturally, the transfer of IPRs will be effective when such rights are in the control of the transferor. Suppose that the token does not inseparably embody such rights. In that case, the person who acquired such rights, even as a result of the initial sale of the token, can dispose of them independently. Then, even if the subsequent agreement suggests an intention to transfer these rights, it will have no legal effect under the principle of *nemo plus iuris*[206] or *nemo dat quod non habet*.[207] As the previous sections show, this is a significant problem. The ability to unbundle these assets is rarely limited.[208]

---

[203] Looking at the 'successes' in this market, this is probably not common knowledge, see Adrienne Westenfeld, 'The Saga of the "Dune" Crypto Bros And Their Very Pricey Mistake Is At Its End' (*Esquire*, 28 July 2022) <https://www.esquire.com/entertainment/books/a38815538/dune-crypto-nft-sale-mistake-explained/> accessed 23 May 2023.

[204] 'Terms of Use' (*Decentraland*) <https://decentraland.org/terms/> accessed 23 May 2023.

[205] Mintable, 'Minting a Gasless NFT on Mintable' (*Mintable Editorial*, 22 August 2021) <https://editorial.mintable.com/minting-a-gasless-nft-on-mintable/> accessed 23 May 2023.

Second, many jurisdictions have specific formal and legal requirements for the effective transfer of such rights. According to Polish law, a copyright transfer agreement must be in writing and signed.[209] The same is true in the U.S. and UK.[210] The difference is that Polish law is very strict and requires such signature to be handwritten. A facsimile or scan is not considered handwritten.[211] Alternatively, the equivalent of a handwritten signature is a qualified electronic signature.[212] Such a strict perception of written form requirements does not correspond to the nature of crypto-asset trading. U.S. law imposes much lower requirements. In the E-Sign Act,[213] electronic signature means 'an electronic sound, symbol, or process, attached to or logically associated with a contract or other record and executed or adopted by a person with the intent to sign the record.'[214] Therefore, the U.S. Court of Appeals for the Fourth Circuit held that accepting the terms of service by clicking the 'yes' button was sufficient to satisfy the requirement for a copyright transfer.[215] Similarly, in UK, under the Interpretation Act 1978, writing means any way of representing words in a visible form.[216] The requirements for recognizing signs as signatures were derived from the *Caton v Caton* case, in which it was held that the intent to use certain signs as a signature was decisive.[217] Thus, an email message,[218] including an automatically added email footer,[219] a checkbox,[220] and a SWIFT message header,[221] meets the requirements.

---

[206] D 50 17 54 Ulpian *On the edict* Bk 46.

[207] Rahmatian (n 4) 220 et seq.

[208] See Part D.

[209] Ustawa o prawie autorskim i prawach pokrewnych 1994 (Polish Copyright Act), art 53. For same requirement for an exclusive license, see art 67(5).

[210] 17 USC § 204(a), § 101; Copyright, Designs and Patents Act 1988, s 90(3), 92(1).

[211] Sąd Najwyższy [Polish Supreme Court], 30 December 1993, III CZP 146/93.

[212] Regulation (EU) No 910/2014 of the European Parliament and of the Council of 23 July 2014 on electronic identification and trust services for electronic transactions in the internal market and repealing Directive 1999/93/EC [2014] OJ L257/73, art 25(2).

[213] 15 USC § 7001 et seq.

In several European countries, copyright agreements must also specify the manner of exploitation of the work to which they apply.[222] However, usually, the regulations provide that, if the contract does not specify the manner of use of the work, such manner is assumed to be based on the nature and purpose of the work.[223] The US Court of Appeals correctly held in *Radio Television Espanola SA v New World Entertainment Ltd* that a copyright assignment agreement does not require 'magic words' but only a sufficient expression of intent to enter into the agreement.[224] This statement deserves full approval. A law is not a set of magic spells that must be uttered to cause legal effects. An agreement is not a Magna Carta; even a one-line statement will be sufficient.[225] Based on the above statement, in *Johnson v Storix Inc*, it was held that it was not even necessary to use the term copyright.[226]

---

[214] 15 USC § 7006(5).

[215] *Metro Reg'l Info Sys Inc v Am Home Realty Network Inc* 722 F3d 591 (4th Cir 2013).

[216] Interpretation Act 1978, sch 1.

[217] *Caton v Caton* (1867) LR 2 HL 127; see *Mehta v J Pereira Fernandes SA* [2006] EWHC 813 (Ch).

[218] *Golden Ocean Group v Salgaocar* [2012] EWCA Civ 265. Note that the case involved the use of the name 'Guy' as a signature, which the court found sufficient to authenticate the message from broker Guy Hindley.

[219] *Neocleous v Rees* [2019] EWHC 2462 (Ch). The above decision is not binding on other courts, but seems to be valid, see The Law Commission, 'Law Com No 386: Electronic Execution of Documents' (Law Commision 9 March 2019).

[220] *Bassano v Toft* [2014] EWHC 377 (QB) [43]-[44].

[221] *WS Tankship II BV v The Kwangju Bank Ltd* [2011] EWHC 3103 (Comm) [155].

[222] Eg Ustawa o prawie autorskim i prawach pokrewnych 1994 (Polish Copyright Act), art 41(2); Ley de Propiedad Intelectual, regularizando, aclarando y armonizando las disposiciones legales vigentes 1996 (Spanish Intellectual Property Act), art 43(1); Gesetz über Urheberrecht und verwandte Schutzrechte (Urheberrechtsgesetz) (UrhG) 1965 (German Copyright Act), para 31(1).

Given the above, using an appropriate private key hash as a signature in the context of circumstances expressing an intent to transfer copyright should be sufficient to meet the formal requirements of US and UK law. This is also important for EU entities who may mistakenly believe that compliance with their national law is always required. Meanwhile, for non-consumer transactions, it is sufficient that at least one of the parties is from the US or UK to satisfy the formal requirements pursuant to Article 11 of the Rome I Regulation.[227]

Suppose it turns out that there are no grounds for an effective transfer of copyright. In that case, one should investigate whether the creators provided a corresponding license. For example, the Bored Ape Yacht Club project grants the NFT owner a worldwide, royalty-free license to use, copy, and display the purchased works, together with any extensions the NFT owner creates or uses, including derivative works.[228]

---

[223] Ustawa o prawie autorskim i prawach pokrewnych (Polish Copyright Act) 1994, art 49(1); Ley de Propiedad Intelectual, regularizando, aclarando y armonizando las disposiciones legales vigentes 1996 (Spanish Intellectual Property Act), art 43(2); Gesetz über Urheberrecht und verwandte Schutzrechte (Urheberrechtsgesetz) (UrhG) 1965 (German Copyright Act), para 31(5).

[224] *Radio Television Espanola SA v New World Entertainment Ltd* 183 F3d 922, 927 (9th Cir 1999).

[225] *Effects Associates Inc v Cohen* 908 F2d 555, 557 (9th Cir 1990).

[226] *Johnson v Storix Inc*, No 16-55439, 1-4 (9th Cir Dec 19, 2017).

[227] Regulation (EC) No 593/2008 of the European Parliament and of the Council of 17 June 2008 on the law applicable to contractual obligations (Rome I) [2008] OJ L177/6.

In the absence of relevant legal documents, many circumstances may also suggest granting at least an implied, nonexclusive license.[229] By acquiring an NFT, the user expects a license to use the works associated with it for their intended purpose. For example, the jurisprudence of the courts in Poland indicates that it is sufficient that the parties in some way intended to grant such a license.[230] In the US, implied licenses have been found in many relationships, particularly in justifying the rights of tattoo holders to use their tattoos. In *Solid Oak Sketches LLC v 2K Games Inc*, the court held that the tattoo artists granted their customers nonexclusive licenses to use their tattoos as part of their image.[231] On the other hand, in *Alexander v Take-Two Interactive Software Inc*, the tattoo artist testified that he had never given any consent to any customer, so it was concluded that there were insufficient grounds for recognizing the existence of an implied license.[232] Such argumentation is questionable because it leads to a determination of intent from the perspective of one party to the transaction rather than taking the objective standard of a third party. In my opinion, for digital assets, it is reasonable to consider, based on the *United States v Univis Lens Co* case, that any situation in which a particular good is acquired is accompanied by the right to use it.[233] Similarly, Lord Hatherley LC stated that a person who buys an article expects control over it. Thus, if a seller does not intend to give the buyer a license to sell or use the product, they must have an express agreement to that effect.[234]

---

[228] 'Terms & Conditions' (*BAYC*) <https://boredapeyachtclub.com/#/terms> accessed 23 May 2023. At the same time, para (i) states that by purchasing NFTs, the buyer also becomes the owner of the 'underlying Bored Ape, the Art, completely,' which could also argue that the intention of the parties was nevertheless to transfer copyright.

[229] *IAE Inc v Shaver* 74 F3d 768, 775 (7th Cir 1996) (holding that the written form requirement applies only to the transfer of copyright and exclusive licenses; nonexclusive licenses may even be implied). If a nonexclusive license is granted for consideration, it is irrevocable by default, see *Asset Marketing v Gagnon* 542 F3d 748, 758 (9th Cir 2008); *Lulirama Ltd v Axcess Broadcast Services* 128 F3d 872, 882 (5th Cir 1997).

[230] Sąd Najwyższy [Polish Supreme Court], 14 September 2005, III CK 124/05.

[231] *Solid Oak Sketches LLC v 2K Games Inc* 449 F Supp 3d 333, 345-46 (SDNY 2020).

[232] *Alexander v Take-Two Interactive Software Inc* 489 F Supp 3d 812, 820 (SD Ill 2020). At the same time, the court rejected defenses based on fair use and a de minimis use, which could also be valuable defenses in crypto-asset cases, see ibid 820-823.

This problem concerning implied rights also exists in some EU regulations. Hence, Article 5(1) of Directive 2009/24/EC[235] stipulates that a lawful acquirer of a computer program does not need to obtain the rightsholder's consent to reproduce and modify the program as necessary to use it in accordance with its intended purpose.[236] A similar provision is found in Article 5(1) of Directive 2001/29/EC,[237] which authorizes temporary acts of reproduction that are transient or incidental and are integral and essential parts of a technological process, and whose sole purpose is to enable (a) transmission over a network between third parties by an intermediary or (b) lawful use, as long as these acts have no independent economic value on their own. This means that, in principle, a license for using a particular digital asset is not always necessary in the EU. In particular, if NFT buyers or users browsing a marketplace want to view digital content on their browsers, they can invoke this exception.[238] The exception may not apply, however, to the display of such content to guests in a restaurant. This is because such further activity may already constitute communication to the public, to which the exception does not apply. Undoubtedly, the public's expectations will justify granting some right to communicate these works to the public by treating such copies as if they were the property of the purchaser of the token, just like a painting on canvas. This leads to another problem regarding further disposals of the token.

---

[233] *United States v Univis Lens Co* 316 US 241, 249 (1942).

[234] *Betts v Willmott* (1871) 6 LR Ch App 239 [245].

[235] Directive 2009/24/EC of the European Parliament and of the Council of 23 April 2009 on the legal protection of computer programs (Codified version) (Text with EEA relevance) [2009] OJ L111/16.

[236] The provision begins with the phrase 'In the absence of specific contractual provisions,' which is a common argument that parties can contractually exclude its application, see Case C-13/20 *Top System SA v État belge* [2021] ECLI:EU:C:2021:811, paras 64-68. However, this provision is intended to ensure that even if the license says otherwise, the law guarantees that the lawful acquirer can use a copy of the program for its intended purpose. Therefore, the reference is to the 'absence' of relevant provisions, not 'unless otherwise provided by contract.'

[237] Directive 2001/24/EC of the European Parliament and of the Council of 4 April 2001 on the reorganization and winding up of credit institutions [2001] OJ L125/15, art 5(1).

One of the exclusive rights of creators is the distribution right. The concept that protects secondary purchasers is referred to as the 'first sale doctrine' or 'exhaustion.' This generally means that creators cannot prohibit a legal purchaser of copies of their work from further using such copies. Unlike most fundamental rights of creators, this rule is not subject to international harmonization. Thus, the scope of this concept varies depending on the legal system and the type of work.

In the US, the first sale grants purchasers the right to display, sell, and otherwise dispose of their purchased copy.[239] However, renting, leasing, lending, or similar acts of temporary use are generally not permitted for computer programs and sound recordings (including musical works contained in such sound recordings) if done for profit, with certain exceptions for embedded computer programs.[240] US law also adopts the principle of international exhaustion, which applies to copies made outside the country and imported from other markets.[241] In the EU, the first sale or other transfer of ownership of the original or copies of the work with the rightsholder's consent exhausts the right to distribute such copies, except for further rental for profit and for non-commercial lending through establishments accessible to the public.[242] In addition, the exhaustion is separately regulated for computer programs, where it generally applies only to the first sale of a copy of the program and provides an exception to exhaustion for the right of commercial rental.[243] The EU adopts the principle that exhaustion applies only to copies first placed on the market in the European Economic Area and prevents EU member states from adopting different exhaustion rules.[244] It should go without saying that in the current age of the global digital economy, adopting any rule different from international exhaustion is inconsistent with society's expectations for trading digital assets.[245] The UK currently appears to be upholding European solutions to the Brexit Act provisions. For now, the government has decided not to make any changes.[246] Leaving aside further details, classically, exhaustion doctrine is perceived in each case through tangible copies of the work. Therefore, a problem has arisen regarding how to apply it to digital copies of works.

---

[238] See Case C-360/13 *Public Relations Consultants Association Ltd v Newspaper Licensing Agency Ltd* [2014] ECLI:EU:C:2014:1195, paras 25-63 (ruling that copies made on a user's screen and cached while browsing the site do not infringe copyright).

[239] However, in the case of audiovisual works, it does not include the right to publicly perform such works. Therefore, the person who publicly plays such a work may be committing copyright infringement.

[240] 17 USC § 109(b)(1)(A)–(B).

---

[241] *Kirtsaeng v John Wiley & Sons Inc* 568 US 519 (2013).

[242] Directive 2001/29/EC, art 4(2).

[243] Directive 2009/24/EC, art 4(2). See recital 12, which states that the rental right does not include public lending, which is outside the scope of the Directive.

[244] Case C-479/04 *Laserdisken ApS v Kulturministeriet* [2006] ECLI:EU:C:2001:577, paras 17-27.

[245] See Enrico Bonadio, 'Parallel Imports in a Global Market: Should a Generalised International Exhaustion Be the Next Step?' (2011) 33 EIPR 153.

[246] Intellectual Property Office, 'UK's Future Exhaustion of Intellectual Property Rights Regime' (*GOV.UK*) <https://www.gov.uk/government/consultations/uks-future-exhaustion-of-intellectual-property-rights-regime> accessed 23 May 2023.

The first sale doctrine does not apply to digital copies in the US.[247] According to the Digital Millennium Copyright Act (DMCA) Section 104 Report, allowing people to transfer digital content would be unfeasible due to the possibility of cheating. The technology at the time made it impossible to implement a forward-and-delete mechanism. In addition, digital files do not degrade over time, so there is an increased risk of piracy.[248] Consequently, in the *Capitol Records LLC v ReDigi Inc* case, both courts held that the digital resale of music files purchased from the iTunes Store through the ReDigi service was not permissible. This is because it constitutes an infringement of the reproduction right, since digital data transfer from one medium to another creates an unauthorized copy.[249]

In EU law, CJEU case law is decisive for recognizing digital exhaustion. The CJEU's decisions interpreting EU law are binding on national courts in all similar cases. The first dispute was decided in *UsedSoft GmbH v Oracle International Corp*.[250] The case involved Oracle's copyright infringement lawsuit over the sale of used licenses for their database software distributed chiefly through digital channels. The language of the license stated that in exchange for payment for services, the user received a non-transferable right to use everything Oracle developed and provided under the agreement for internal business purposes only, free of charge, for an unlimited period.[251] Meanwhile, UsedSoft decided to buy used licenses for Oracle software from the original buyers and resell them on the market. However, they did not give their customers copies of the programs but only referred them to Oracle's website to download the software themselves or encouraged them to duplicate copies they already had.[252] The CJEU stated that it is sufficient to consider that there has been a transaction of putting the product on the market by transferring ownership to the purchaser in exchange for a price corresponding to the economic value of the copy of the work.[253] The fact that the copy was made available to the purchaser through digital distribution is insignificant.[254] Entering into a license agreement and downloading a copy of a program are two inseparable transactions.[255] The ruling has been criticized as an intrusion by the CJEU into EU member states' authority to regulate property rights.[256] Such a position seems unjustified, since the decision only concerned exhaustion, which is a copyright construct that serves only to clarify the rules of secondary trade in copies of works.

---

[247] Sarah Reis, 'Toward a "Digital Transfer Doctrine"? The First Sale Doctrine in the Digital Era' (2015) 109 Northwestern University Law Review 173, 183–85.

[248] U.S. Copyright Office, 'DMCA Section 104 Report' (US Copyright Office August 2001) 78–101.

[249] *Capitol Records LLC v ReDigi Inc* 934 F Supp 2d 640, 654–56 (SDNY 2013); *Capitol Records LLC v ReDigi Inc* 910 F3d 649, 655–60 (2d Cir 2018). However, the Court of Appeals considered only the argument regarding the fact of unlawful reproduction without addressing in detail the question as to whether exhaustion could apply not only to the 'original' copy but also to the copy created as a result of the transfer, see ibid 656.

[250] *UsedSoft GmbH* (n 59).

Due to the separate regulation of the exhaustion for works other than computer programs, doubts have arisen as to whether the above considerations apply in the same way to other goods, such as e-books.[257] The dispute over e-books was partially resolved in the 2019

---

[251] ibid para 23.

[252] ibid para 26.

[253] ibid para 35-72.

[254] ibid para 50-63.

[255] ibid para 43-49.

[256] Eg Péter Mezei, 'Digital First Sale Doctrine Ante Portas – Exhaustion in the Online Environment' (2015) 6 JIPITEC 23 (summarizes much of criticism).

[257] This is also relevant to the software, since most contain both literary works (programs) and various types of audiovisual content. This consideration led CJEU to recognize that video games are protected as such, not as

Tom Kabinet case.[258] At the time, the court held that an e-book is not a computer program, so the provisions of Directive 2009/24/EC shall not apply to it.[259] In addition, the right of distribution and its exhaustion from Directive 2001/29/EC apply only to tangible copies of works, and thus, do not apply to e-books.[260] Therefore, CJEU concluded that reselling used e-books through a website constitutes an act of communication to the public, even if a copy of the work is given only to one user, since the potential audience is unknown.[261] It seems that such a position, in contrast to that in the UsedSoft case, does not account for the economic significance of the transaction and the technical mode of operation. The court merely stated that, from an economic and functional point of view, an e-book and a physical book are not equivalent.[262] Digital copies are not degraded with use, and replacing them does not require additional effort or cost. Therefore, a parallel market for used digital copies risks infringement on the interests of rightsholders more than in the case of physical copies.[263] A brief study of the current situation shows that the same comments can be made about computer programs and their tangible media. Nevertheless, the court created a market for used digital copies of computer programs in the UsedSoft ruling. It is also impossible not to mention that the Tom Kabinet judgment clearly contradicts an earlier CJEU judgment on the public lending of e-

---

computer programs, but under Directive 2001/29/EC, see Case C-355/12 *Nintendo Co Ltd v PC Box Srl and 9Net Srl* [2014] ECLI:EU:C:2014:25, para 23.

[258] Case C-263/18 *Nederlands Uitgeversverbond and Groep Algemene Uitgevers v Tom Kabinet Internet BV* [2019] ECLI:EU:C:2019:1111.

[259] ibid paras 53-59.

[260] ibid paras 36-52.

[261] ibid paras 60-72.

[262] ibid para 58.

[263] ibid.

books.[264] That judgment held that EU member states may condition the application of exceptions to such a right on the sale of e-books in a way that leads to exhaustion under Directive 2001/29/EC.[265] However, since there can be no exhaustion of e-books, such an exception will never apply.

Also problematic for the proper perception of NFT transactions is the correct distinction between a copy of a work and a means of accessing it. In the case of *Disney Enterprises Inc v Redbox Automated Retail LLC*, it was held that a copy of a work is different from a code that allows its downloading from a vendor's website.[266] The issue was that Redbox had acquired bundles of movies with codes for downloading additional content and began selling them separately to its customers. Disney changed the terms of the licensing agreement so that downloading additional content was allowed only for personal use and only if one owned the original disc. Therefore, Disney sued Redbox, believing that the latter indirectly contributed to copyright infringement by encouraging its customers to make illegal copies. In the case, the court rejected Redbox's defense based on the first sale doctrine since no copy existed yet at the time of the sale, and the sale of the code is not equivalent to the sale of a copy of the work.[267]

Tokens, by their nature, do not constitute copies of a work but only may, in certain circumstances, represent the right to use such work. Consequently, the courts may find that an NFT purchaser is not protected by the exhaustion but is an indirect infringer encouraging the creation of illegal copies by subsequent token purchasers. However, such an observation is incompatible with the previously described intention of the parties. It also looks quite radical and far-fetched. Using any digital distribution mechanism, primarily through access codes, would lead to the recognition that there would be no exhaustion of a copy of the work, as this copy would only come into existence at the time of downloading. Therefore, the approach outlined in the Redbox ruling should not be considered appropriate. We should look for a relatively objective intention of the parties manifested in their particular behavior. Exhaustion should be viewed through its function of ensuring the free movement of goods, balancing the interests of creators and users.

---

[264] Case C-174/15 *Vereniging Openbare Bibliotheken v Stichting Leenrecht* [2016] ECLI:EU:C:2016:856, para 15.

[265] ibid paras 55-65.

[266] *Disney Enterprises Inc v Redbox Automated Retail LLC* 336 F Supp 3d 1146 (CD Cal 2018).

[267] ibid 1156–57.

I believe that the above positions are incompatible with the current social and market situation. Exhaustion should be extended to the digital environment and applied in full to transactions that are the economic equivalent of acquiring copy of a work. In doing so, it is crucial to properly distinguish the various purposes of transactions. Exhaustion should not occur when the subject of the agreement is only a service (for example, streaming, software as a service, or social media access).[268] However, if the primary purpose of the transaction is to provide the vendor with a benefit that can be considered payment for a sale, the transaction should be treated as such. The language of the contract between the parties should not be determinative and should be evaluated in this context from the perspective of an objective third party. In this respect, the right of communication to the public should be considered as the right to take actions that result in an indeterminate and substantial number of potential recipients having the opportunity to become acquainted with the work. However, this should not per se include situations in which the recipient does not have the opportunity to use a particular form of the work in accordance with its economic purpose.[269] All other activities should be treated as other exclusive rights, including any transfer of a work to a single recipient as a distribution and reproduction right.[270] However, if a work is reproduced during digital resale because of the state of the art, such reproductions, which are incidental to distribution, should not constitute copyright infringement.[271] Other approaches encourage the creation of service monopolies and the offering of all products exclusively under the Everything as a Service (XaaS) model, ultimately making ownership a relic of the past.[272]

---

[268] The first sentence of recital 29 of Directive 2001/29 states that exhaustion does not apply to services, in particular online services. However, this does not mean that every act performed online is a service and not a distribution. Recital 25 emphasizes that communication to the public refers specifically to on-demand transmissions, but says nothing about a single transmission of a copy of a work.

When discussing IP issues, it is impossible not to mention the problem of protecting machine-generated creations, considering the popular so-called 'generative art NFTs' whose final appearance is pseudo-random. In such creations, it is challenging to determine the buyer's IP, as the buyer's role is usually not even limited to selecting variants from the available pool. Instead, the buyer's role is akin to triggering a lottery machine. Perhaps such results are the IP of the smart contract developer? It can be said that at the stage of creating such a program, it is likely to predict all possible variants of how the generated content may look like. However, IP law does not protect the ideas but only their form of expression. Thus, each image must be recognized as having been predicted and programmed in advance as a certain outcome that represents someone's creativity. However, such a situation would be relatively rare and inconsistent with how real generative systems work.[273] For related rights, such as phonogram producers' rights, the situation is different. They do not require creative input and output, so they can be protected even if they are created by machines. However, such a distinction no longer seems justified, especially since related rights are not as widely recognized as copyright.[274] There is also no general rule that discriminates against certain types of works, so there should be no such rule for different fixations. Therefore, a new regime is needed.

---

[269] Eg streaming GUIs or even entire digital worlds (including the metaverse) if the recipient of such a transmission cannot actively use their functionality; cf Case C-393/09 *Bezpečnostní softwarová asociace – Svaz softwarové ochrany v Ministerstvo kultury* [2010] ECLI:EU:C:2010:816, paras 52-58 (stating that television broadcasting of a GUI does not constitute communication to the public). However, this exception should perhaps be based on fair use rather than an exclusion from the scope of copyright.

[270] See Case C-637/19 *BY v CX* [2020] ECLI:EU:C:2020:863, para 29; Case C-325/14 *SBS Belgium NV v Belgische Vereniging van Auteurs, Componisten en Uitgevers (SABAM)* [2015] ECLI:EU:C:2015:764, paras 23-24.

[271] Similar to the wording of Directive 2009/24/EC, art 5(1); Directive 2001/29/EC, art 5(1); 17 USC § 117(a)(1).

[272] See monopolization of databases in the EU that are not protected by *sui generis* or copyright law. The provider of such a database may contractually specify any conditions for using it, see Case C-30/14 *Ryanair Ltd v PR Aviation BV* [2015] ECLI:EU:C:2015:10.

The solution may be the right to first fixations of information, especially in digital form. This right should operate independently of copyright and other IPRs. What is needed is a regime that simultaneously protects the results of machines, including artificial intelligence, and does not interfere with the protection of digital data as objects of property rights, but is a necessary complement to them. Perhaps it could even be a regime that would render all other related rights superfluous, covering within its scope any first fixation.[275] I propose that the person who directly causes the creation of such a fixation be considered the holder of the mentioned right (so, as a rule, the tool's user, not its manufacturer).[276] Nevertheless, effective implementation of this approach requires the absolute recognition of digital exhaustion, regardless of where copies are made.

---

[273] See U.S. Copyright Office, 'Copyright Registration Guidance: Works Containing Material Generated by Artificial Intelligence' (16 March 2023) (stating that AI contributions cannot be protected because they lack human authorship, but this does not apply when humans select or arrange AI-generated material).

[274] Therefore, the unification of the protection of these creations under copyright law in the US may raise some doubts. Sometimes, however, it is possible to benefit from the doctrine of misappropriation.

[275] Perhaps the condition of originality, which has always been a problem in distinguishing unprotected ideas from protected expression, should be abandoned. Possibly such a right to first fixation of information could also replace copyright in the future.

## 7. Conclusion

The above considerations show an emerging need to carve out a legal regime for the ownership of tokens, which are more broadly digital data. I have long argued that it is necessary to universally recognize digital data as objects of property rights. This does not mean that all data will now be subject to such a monopoly, just as not all elementary particles are subject to such rights (such as free air or open sea). On a case-by-case basis, it must be determined whether certain data can stand alone as objects of circulation.

A particular approach in this regard has been proposed by the Law Commission—to introduce data objects as the third category of property objects.[277] Property objects would be considered data objects if they meet the following three conditions: (1) they consist of data represented in an electronic medium; (2) their existence is independent of persons and the legal system; and (3) they are rivalrous. However, according to the authors, these criteria will not be met by digital files, digital records,[278] email accounts, certain in-game items,[279] domain names,[280] and carbon emissions trading schemes.[281] The authors currently include only crypto tokens in this category.

---

[276] See provisions for computer-generated works in Indian Copyright Act 1957, s 2(d)(vi); cf Copyright, Designs and Patents Act 1988, s 90(3); Irish Copyright and Related Rights Act 2000, s 21(f); New Zealand Copyright Act 1994, s 5(2)(a). However, since any work must be the product of a human mind, such protection will be ineffective in most of these jurisdictions.

[277] The Law Commission (n 84) 77 et seq.

[278] ibid 109–23.

[279] ibid 124–38.

Clearly, such a proposal is very myopic. This is confirmed by the arguments in the discussion above that most of these goods are capable of existing as independent objects of trade. The main problem with the Law Commission's proposal is the second criterion, which excludes email accounts, domain names, and in-game items. In economic terms, these are always particular goods that result from the existence of an entry in a specific database. The difference is that their records are centralized, while crypto-assets are distributed. Therefore, crypto-assets represent a certain value resulting from the fact that their database can be maintained by many. Nevertheless, both assets are created only by a certain social consensus to assign a value to these records, which can be traded.[282]

There should be no doubt about the right to dispose of digital files. Some people say that a file cannot be transferred between devices in such a way that it is completely removed from the original device during the process. File transfer creates a new file. According to some people, this makes it impossible to consider digital files as objects of exclusive rights. How then should we see plants growing from a mere seed to a large tree? In principle, the growth of a plant can be compared to copying as the volume of the plant increases. No one denies that such 'copied' elements of a seed—its trunk, bark, branches, and leaves—can be things. Similarly, the same could be said of plant propagation. Copying files can be seen in the same way. A property right is not only the right to exclusivity, but first and foremost the right to use. The natural way to use a file is to copy it. If this is a problem for lawyers, the moment cloning comes into public use, one must predict the doom of legal systems. Just as the invention of the printing press doomed writers, and when widespread access to computers and copying technology led to the death of copyright.

---

[280] ibid 139–45.

[281] ibid 146–55.

[282] Similar comments were made by several organizations in response to the consultation, such as the Electronic Money Association, Queen Mary University of London and Meta, which properly emphasized that data regulation should not discriminate based on the underlying technology, see The Law Commission, 'Digital Assets: Responses to Consultation' <https://s3-eu-west-2.amazonaws.com/lawcom-prod-storage-11jsxou24uy7q/uploads/2022/07/Digital-assets-collated-consultation-responses.pdf> accessed 23 July 2023.

Nor should the criterion for protection be the ability of one person to destroy a particular good of another. As with material objects, there is no invincible force that prevents others from damaging property. That is why the law provides for an obligation to compensate for damage. In practice, the law is only a specific abstract construct intended to reflect social relations. The law always plays a secondary role in society. It should ensure the protection of universally recognized values. Since society expects the protection of the crypto-assets, an approach that limits their protection is completely unjustified.

In addition to issues related to property rights in the strict sense, it is important to change the approach to IPRs by recognizing the need to create new rights to first fixations of information. It is also necessary to change the perception of digital exhaustion. In legal systems based on conservative notions of written form, in particular for copyright transfer agreements, signature requirements need to be aligned with those in common law systems. This is also important for the protection of EU consumers wishing to acquire intellectual property rights, since Article 11(4) of the Rome I states that the law applicable to the form of the contract is the law of the country of the consumer's habitual residence, so the consumer may not acquire any rights. There is also a clear need to create harmonized standards for the identification of securities and the rules for trading in them.

These problems are not new, they have existed for many years, especially in the context of virtual goods, but with the exception of a few researchers, no one saw the need for change.

Today, however, the issue has become particularly important because of the scale of digital distribution, the high value of digital assets, and the vision of a new version of the Internet based on the metaverse concept. As Fryderk Zoll Jr wrote: 'The higher the level of legal development is, the more goods the individual can master.'[283]

---

[283] Fryderyk Zoll Jr, *Prawo Cywilne Opracowanie Głównie Na Podstawie Przepisów Obowiązujących w Małopolsce, t. II, Prawa Rzeczowe i Rzeczowym Podobne* (Third, 1931) 178.